\documentclass[preprintnumbers]{revtex4}
\usepackage{eurosym}
\usepackage{amsmath}
\usepackage{amssymb}
\usepackage{graphicx}
\usepackage{color}

\begin{document}

\date{\today }
\title{Trions in two-dimensional monolayers within the hyperspherical
harmonics method. Application to transition metal dichalcogenides}
\author{Roman Ya. Kezerashvili$^{1,2}$, Shalva M.Tsiklauri$^{3}$, and Andrew
Dublin$^{1,2}$}
\affiliation{\mbox{$^{1}$New York
City College of Technology, The City University of New York, USA} \\
$^{2}$The Graduate School and University Center, The City University of New
York, USA\\
$^{3}$Borough of Manhattan Community College, The City University of New
York, USA}

\begin{abstract}
We develop the theoretical formalism and study the formation of valley
trions in two-dimensional monolayers within the framework of a
nonrelativistic potential model using the method of hyperspherical harmonics
(HH) in four-dimensional space. We present the solution of the three-body
Schr\"{o}dinger equation with the Rytova-Keldysh (RK) potential by expanding
the wave function of a trion in terms of the HH. 

We consider a long-range approximation when the RK potential is approximated
by the Coulomb potential and a short-range limit when this potential is
approximated by the logarithmic potential. In a diagonal approximation, the
coupled system of differential equations for the hyperradial functions is
decoupled in both limits. Our approach yields the analytical solution for
binding energy and wave function of trions in the diagonal approximation for
these two limiting cases: the Coulomb and logarithmic potentials. We obtain
exact analytical expressions for eigenvalues and eigenfunctions for
negatively and positively charged trions. The corresponding eigenvalues can
be considered as the lower and upper limits for the trions binding energies.

We apply the proposed theoretical approach to describe trions in transition
metal dichalcogenides (TMDC) and address the energy difference between the
binding energies of $X^{-}$ and $X^{+}$ in TMDC. Results of numerical
calculations for the ground state energies with the RK potential are in good
agreement with similar calculations and in reasonable agreement with
experimental measurements of trion binding energies.
\end{abstract}

\maketitle


\section{Introduction}

\bigskip In the last two decades, the condensed matter community has
benefited from substantial progress in the understanding of excitons in
band-gap two-dimensional (2D) semiconductors. Excitons are the most
rudimentary bound quasiparticles formed by the electrostatic attraction
between a positive valence-band hole and a conduction-band electron. Due to
reduced dimensionality, diminished dielectric screening, and large charged
carrier masses, the binding energies of excitons are quite large, with some
reported energies as large as 0.5 eV \cite{Kormanyos2015}.

The idea of the existence of a trion that is a bound state of an exciton
with another charged carrier, which can either be an electron ($X^{-}$
trion) or a hole ($X^{+}$ trion) was first suggested by Lampert over 60
years ago \cite{Lampert58}. This idea gave rise to many theoretical and
experimental studies of trions in bulk materials and quantum-well systems
from the 1960s to the 2000s. Only 35 years later, negatively charged trions
were first observed in CdTe quantum-well system \cite{Kheng1993}. The
positively charged trions were discovered three years later, in 1996 \cite%
{Finkelstein1996}. Theoretical calculations have shown that the binding
energies of trions are very small in bulk materials \cite{FilikhinKez}, but
they are substantially enhanced in structures of reduced dimensionality.

Since 2013, when trions were observed in two-dimensional MoS$_{2}$
monolayers \cite{MoS23Heinz}, experimental and theoretical interests in
trions in 2D materials have increased considerably. Reduced dielectric
screening in 2D materials leads to greatly enhanced trion binding energies,
whose signatures appear frequently in the optical spectra of monolayer
transition metal dichalcogenides (TMDCs). TMDCs constitute a unique subgroup
of crystalline, two-dimensional band-gap semiconductors whose compounds have
a chemical formula MX$_{2}$, where M and X denote a transition metal (Mo or
W) and a chalcogenide (S, Se, or Te), respectively. Different experimental
groups \cite{MoS23Heinz,MoSe21 Ross,WSe2
Jones,WSe2Wang,MoSe2Singh,Liu,Zhang2014,bZu,Yang,ShangBiexiton,ZhangMS2,WS2Plechinger,Singh2016,substrate,Christopher2017,CourtadeSemina,Volmer2017,Liu2019,Borghardt2020,ExpZinkiewicz2022,Klein2022,Semina2022}
observed and reported the signature of a trion in TMDCs formed from a hole
in the first valence band and electrons originating predominantly from
downwards curved conduction bands of the $K$ and/or $K^{\prime }$ valleys. 

Many studies have been carried out to calculate the binding energies of
excitonic complexes in monolayer TMDCs (see reviews \cite%
{NScRev2015,BekReichman,Durnev2018,Kezerashvili2019,Suris2022}). To study
the physical properties of trions in 2D TMDC, one should know the wave
function and the corresponding energy levels for a three-particle system.
Theoretical studies of $X^{-}$ and $X^{+}$ trions \cite%
{Reichman2013,TimeDepdensity matrix functional
theory,BerkelbachDifMonteCarlo,DenFuncTheoryPIMC,Saxena,VargaNano2015,Ganchev,VargaPRB2016,Szyniszewski,Szyniszewski2,Deilmann2017,Danovich2017,KezFew2017,Efimkin2017,magnetic,Chaves2018,Florian2018,Torche2019,photolum,Fu2019,Varga2020,ZhumagulovTamm,ZhumagulovTamm2,Glazov2020,model,Variational2021,Marsusi2022,Glazov2023,Frederico2023}
have integrated a wide variety of techniques and approaches, yielding
impressively accurate results consistent with experimental data.

Within the framework of the variational method, a microscopic theory of an
effective mass model of trions in TMDC monolayers was developed in Ref. \cite%
{Reichman2013}. The calculated binding energies are in good agreement with
many-body computations based on the Bethe-Salpeter equation. Dark trions
were predicted in Refs. \cite{Deilmann2017,Danovich2017}. More recently,
trions have been studied using the variational method with trial wave
functions constructed from 2D Slater-type orbitals \cite{Variational2021}.
Using the variational method, the trion binding energies and corresponding
wave functions can be calculated efficiently with fairly good accuracy \cite%
{Variational2021,Semina2022}.

The diffusion quantum Monte Carlo method was used for the Mott-Wannier model
of trions in 2D semiconductors \cite%
{BerkelbachDifMonteCarlo,Szyniszewski,Szyniszewski2} and, most recently, for
high-precision statistically exact quantum Monte Carlo calculations of the
binding energies of 3D excitonic complexes \cite{Marsusi2022}. Researchers
have also made extensive use of the path-integral Monte Carlo method \cite%
{DenFuncTheoryPIMC,Saxena} to study the effects of dielectric screening on
trion binding energies, often incorporating stochastic variational
techniques to improve ground-state energy estimates. Both the diffusion and
the path-integral Monte Carlo methods have proven to be very effective in
numerically solving the time-independent Schr\"{o}dinger equation.

The stochastic variational method is applied to study the formation of
trions within TMDC monolayers using a correlated Gaussian basis \cite%
{VargaNano2015,VargaPRB2016} to investigate binding energies and their
dependence on both the effective screening length and the electron-hole
effective mass ratio. Stochastic variational methods have also been utilized
to examine the effects of external magnetic fields and phonon recombination
processes on excitonic binding energies \cite{magnetic}. Within the
framework of the stochastic variational approach using the complex scaling
and the stabilization method, it has been shown that there are narrow
resonance states in two-dimensional three-particle systems of electrons and
holes interacting via a screened Coulomb interaction \cite{Varga2020}.

In Ref. \cite{Ganchev} the authors evaluate binding energies of $X^{\pm }$
trions by mapping the three-body problem with three logarithmically
interacting particles in two dimensions onto one particle in a
three-dimensional potential treated with the boundary-matching-matrix
method. The trion states are calculated by a direct diagonalization of the
three-particle Hamiltonian within the Tamm--Dancoff approximation \cite%
{ZhumagulovTamm,ZhumagulovTamm2}.

Trions in TMDC monolayers have been investigated using both multi-band and
single-band models, and the finite element method has been applied to both
models \cite{model}. In the framework of the multi-band model, the authors 
\cite{model} constructed the excitonic Hamiltonian in the product base of
the single-particle states at the conduction and valence band edges and
solved the energy eigenvalue equation using the finite element as well as
the stochastic variational \cite{VargaPRB2016,Suzuki} methods.

The structural complexity of TMDCs gives rise to striking energetic
properties. The periodicity of TMDC crystalline lattices permits the
occupation of not only multiple charge and spin configurations, but also
degenerate energy levels in the valence and conduction bands \cite%
{light-valley}. These bands are separated (in momentum space) by $K$ and $%
K^{\prime }$ electron valleys \cite{substrate}; the energy degeneracy at the
valley boundary appears to be a direct result and manifestation of
spin-orbit coupling and spin-splitting \cite{valley,CourtadeSemina,revealing}%
. Researchers have posited that trion complexes can exist in different
spin-valley configurations: the singlet and triplet states. The triplet
state is more energetic due to long-range intervalley Coulomb attraction
between the exciton and an additional conducting-band electron \cite{valley}%
. Substrate screening and doping are also believed to be significant
contributors to these higher binding energies \cite{substrate}.
Interestingly, photoluminescence spectra in TMDC monolayers such as WSe$_{2}$
suggest trion sensitivity to temperature \cite{valley,excited-state,Fu2019}:
as the temperature rises, the triplet states are believed to be more densely
occupied than their singlet counterparts \cite{valley}.

Much of our current understanding of exciton transitions and binding
energies has emerged from optical spectroscopy \cite%
{CourtadeSemina,substrate,coupling,Florian2018,revealing}. Photoluminescence
spectra and optical reflectivity measurements of WSe$_{2}$ monolayers
suggest a significant discrepancy between the binding energies of positive
and negative trions \cite{CourtadeSemina}. In the effective mass
approximation, it is hypothesized that short-range electron-hole and
electron-electron Coulomb interactions \cite%
{Glazov2020,coupling,CourtadeSemina} give rise to $X^{-}$ fine-structure
splitting in WSe$_{2}$ monolayers, providing a possible explanation as to
why the $X^{-}$ binding energy is 
larger than that of the positive trion \cite{CourtadeSemina}. Similarly,
optical red-shifts in the absorption spectra of MoS$_{2}$ monolayers suggest
that reductions in the exciton and trion binding energies are attributable,
in large part, to a significant substrate-induced electric screening,
substrate polarization, and doping, all of which are believed to diminish
the Coulomb exchange interactions \cite{substrate}. The theory of high-lying
trions in 2D semiconductors with negative effective mass has been developed
using the variational approach \cite{Glazov2023}. Authors have demonstrated
the key role of the non-parabolicity of the high-lying conduction band
dispersion in the formation of the bound exciton and trion states.

Excitonic systems are many-body systems, and the most systematic approach
requires the use of quantum field theory. However, these excitonic systems
can be well approximated and treated in the framework of few-body physics.
There are different approaches to solving the three-body eigenvalue and
eigenfunction problem in two dimensions for interacting electrons and holes 
\cite{Kezerashvili2019}. The Faddeev equations and hyperspherical harmonic
(HH) methods are commonly used in nuclear and atomic physics to solve the
three-body problem in three-dimensional configuration or momentum spaces.
Both methods are useful in studying three charged particles in a
two-dimensional harmonic well and in calculating 2D trion energy levels, and
both are implemented in Refs. \cite{Braun1,
Braun2,KezFew2017,FilikhinKez,Frederico2023}.

In this paper, we develop a theoretical formalism for the three-body problem
using the method of HH. The three-body problem is solved for one-layered
systems within the effective mass approximation using the Rytova-Keldysh
(RK) potential \cite{Rytova,Keldysh,Rubio}. The HH method allows us to map
the three-body problem in 2D configuration space onto a one-body problem in
four-dimensional (4D) space. Separation of variables is invoked in the
four-dimensional hyperspace to decouple the radial and angular dependence of
the three-body wave function. Specifically, the wave function is expanded in
terms of the basis of angular eigenfunctions of the four-dimensional Laplace
operator. Our approach yields the analytical solution for binding energies
of trions in the diagonal approximation in two limiting cases - the Coulomb
and logarithmic potentials. For the complete solution of the three-body Schr%
\"{o}dinger equation with the effective two-dimensional screened potential 
\cite{Rytova,Keldysh}, we expand the wave functions of three bound particles
in terms of the antisymmetrized hyperspherical harmonics. We then
numerically solve the resulting coupled system of the second order
differential equations. 

This article has two foci and consists of two parts: i. the development of
the technique to study trions in 2D monolayers in the framework of HH, and
ii. the verification of this approach by its application to trions in TMDC.
The article is organized as follows. In Sec. II, we present the theoretical
approach and formalism to study Mott-Wannier trions in monolayer 2D
semiconductors within the framework of the method of hyperspherical
harmonics. We consider a non-relativistic potential model for three
interacting particles and employ the three-body Schr\"{o}dinger equation in
the effective mass approximation. For the solution of the Schr\"{o}dinger
equation with the Rytova-Keldysh potential \cite{Rytova,Keldysh}, we expand
the wave function of three bound particles in terms of the antisymmetrized
hyperspherical harmonics, and obtain the corresponding system of coupled
differential equations for the hyperradial functions. In Sec. III, the
diagonal approximation is considered for the coupled differential equations
for the hyperradial functions for two limiting cases of the RK potential:
the Coulomb and logarithmic potentials. Our approach yields the analytical
solutions for binding energies and wave functions of trions in the diagonal
approximation for both the long-range Coulomb and short-range logarithmic
potentials. In Sec. IV, we apply our approach to study trions in TMDC
monolayers. We present and discuss results for the trions binding energies
obtained by the numerical solution of the coupled differential equations for
the RK potential and the results of calculations with the Coulomb and
logarithmic potentials. Conclusions follow in Sec. V.

\section{Theoretical approach and formalism}

This Section provides an outline of the low-energy model that describes
Mott-Wannier trions in 2D 
monolayers. We present a widely used screened electrostatic interaction
between charged carriers in few-body complexes in two-dimensional materials
and develop the hyperspherical harmonics formalism for trions in 2D
monolayers. 

\subsection{Charge-charge interaction}

Within the effective mass approach, the non-relativistic Mott-Wannier trion
Hamiltonian in a 2D configuration space reads 
\begin{equation}
H=-\frac{\hslash ^{2}}{2}\overset{3}{\underset{i=1}{\sum }}\frac{1}{m_{i}}%
\nabla _{i}^{2}+\overset{3}{\underset{i<j}{\sum }}V_{ij}(\left\vert \mathbf{r%
}_{i}-\mathbf{r}_{j}\right\vert ),  \label{Trion}
\end{equation}%
where $m_{i}$ are the effective masses and $\mathbf{r}_{i}$ are the $i$th
particle Cartesian coordinates in 2D space. We assume only two types of
charge carriers: electrons and holes, with the corresponding effective
masses treated within the band effective mass approximation. In Eq. ~(\ref%
{Trion})\ $V_{ij}(\left\vert \mathbf{r}_{i}-\mathbf{r}_{j}\right\vert )$ is
the screened electromagnetic interaction between $q_{i\text{ }}$ and $q_{j}$
point-like charges in 2D material, which was first derived in Ref. \cite%
{Rytova} and independently obtained by Keldysh \cite{Keldysh}. The obtained
spherical symmetry potential describes the interaction between two charged
particles in a film of finite thickness under the intrinsic assumption that
the screening can be quantified by the dielectric constant of the bulk
material. In other words, the thickness of the film is assumed to be large
with respect to the lattice constant. Thus, the dielectric constant is no
longer a good parameter to quantify the macroscopic screening. In this case,
the screening is best quantified by the 2D polarizability. In Ref. \cite%
{Rubio} the authors provide a strict 2D derivation of the macroscopic
screening and obtain the screened charge-charge interaction, which has the
same functional form as the potential \cite{Rytova,Keldysh}. The effective
charge-charge potential that describes the Coulomb interaction screened by
the polarization of the electron orbitals in the 2D lattice reads 
\begin{equation}
V_{ij}(r)=\frac{\pi kq_{i}q_{j}}{2\epsilon \rho _{0}}\left[ H_{0}\left( 
\frac{r}{\rho _{0}}\right) -Y_{0}\left( \frac{r}{\rho _{0}}\right) \right] .
\label{Keldysh}
\end{equation}%
In Eq.~(\ref{Keldysh}) $r$ denotes the relative distance between the charge
carriers, whose electric charges are $q_{i}$ and $q_{j}$, $k=9\times 10^{9}$
N$\cdot $m$^{2}$/C$^{2}$, $\epsilon $ is the dielectric constant of the
environment that is defined as $\epsilon =(\varepsilon _{1}+\varepsilon
_{2})/2$, where $\varepsilon _{1}$ and $\varepsilon _{2}$ are the dielectric
constants of two materials surrounding the TMDC layer, $\rho _{0\text{ }}$
is the screening length, and $H_{0}\left( \frac{r}{\rho _{0}}\right) $ and $%
Y_{0}\left( \frac{r}{\rho _{0}}\right) $ are the Struve function and Bessel
function of the second kind, respectively. In the case of a freestanding
monolayer, in Ref. \cite{Rubio} it is shown that $\rho _{0}=2\pi \chi $,
where $\chi $ is the polarizability of the $2$D material, which sets the
boundary between two different behaviors of the potential due to a nonlocal
macroscopic screening. Following Refs. \cite{Gradshteyn,Abramowitz,Rubio}
one can obtain that for the long-range, when $\ r>>\rho _{0}$, the potential
(\ref{Keldysh}) has the three-dimensional bare Coulomb tail, while for the
short-range, when $r<<\rho _{0}$, it becomes a logarithmic potential like a
potential of a point charge in two dimensions: 
\begin{equation}
V_{ij}(r\mathbf{)=}\left\{ 
\begin{tabular}{cc}
$\frac{kq_{i}q_{j}}{\epsilon r},\text{ \ \ \ \ \ \ \ \ \ \ \ \ \ \ \ \ \ \ }$
& $\text{when }r>>\rho _{0\text{, }}\text{ Coulomb potential,}$ \\ 
$\frac{kq_{i}q_{j}}{\epsilon \rho _{0\text{ }}}\left[ \ln \left( \frac{r}{%
2\rho _{0}}\right) +\gamma \right] ,$ & $\ \ \ \text{when }r<<\rho _{0\text{,%
}}\text{ Logarithmic potential,}$%
\end{tabular}%
\right.  \label{Limit}
\end{equation}%
where $\gamma $ is the Euler constant. Therefore, the potential (\ref%
{Keldysh}) becomes the standard bare Coulomb potential for $r>>\rho _{0\text{
}}$ and diverges logarithmically for $r<<\rho _{0\text{ }}.$ A crossover
between these two regimes takes place around distance $\rho _{0}$. From Eq. (%
\ref{Limit}) one can conclude that increasing the screening length $\rho _{0%
\text{ }}$leads to a decrease in the short-range interaction strength, which
means that screening is more efficient in highly polarizable 2D materials,
while the long-range interaction strength is unaffected. In Ref. \cite{Rubio}%
, a very good approximation was introduced for the potential (\ref{Keldysh})
in terms of elementary functions that is simpler to use in calculations,
fairly precise in both limits, and accurate for all distances.

\subsection{Hyperspherical harmonics formalism}

\begin{figure}[h]
\centering
\includegraphics[width=12.0cm]{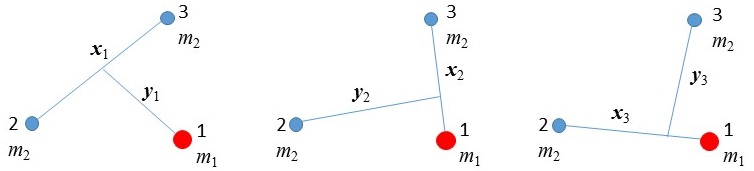}
\caption{(Color online) The partition trees of Jacobi coordinates for $X^{-}$
trion. }
\label{JocCordinate}
\end{figure}

The Schr\"{o}dinger equation for negative and positive trions with the
interaction (\ref{Keldysh}) reads 
\begin{equation}
\left[ -\overset{3}{\underset{i=3}{\sum }}\frac{\hslash ^{2}}{2m_{i}}\nabla
_{i}^{2}+\sum_{i<j}^{3}\frac{\pi kq_{i}q_{j}}{2\epsilon \rho _{0}}\left(
H_{0}(\frac{\left\vert \mathbf{r}_{i}-\mathbf{r}_{j}\right\vert }{\rho _{0}}%
)-Y_{0}(\frac{\left\vert \mathbf{r}_{i}-\mathbf{r}_{j}\right\vert }{\rho _{0}%
})\right) \right] \Psi (\mathbf{r}_{1},\mathbf{r}_{2},\mathbf{r}_{3})=E\Psi (%
\mathbf{r}_{1},\mathbf{r}_{2},\mathbf{r}_{3}),\ \   \label{Schred}
\end{equation}%
where $\nabla _{i}^{2}$ is the 2D Laplace operator and $E$ is the ground or
excited states energy of the three-body system. At the first step one
separates the center-of-mass (c.m.) motion from the relative motion of three
particles through the transformation from the original Cartesian $\mathbf{r}%
_{1},\mathbf{r}_{2},\mathbf{r}_{3}$ coordinates in 2D space to the set of
mass--scaled 2D Jacobi coordinates shown in Fig. \ref{JocCordinate}. We have
three equivalent sets of Jacobi coordinates ($i\neq j=1,2,3$). For the
partition $i$, the mass--scaled Jacobi coordinates are as follows 
\begin{eqnarray}
\mathbf{x}_{i} &=&\sqrt{\frac{m_{j}m_{k}}{(m_{j}+m_{k})\mu }}(\mathbf{r}_{j}-%
\mathbf{r}_{k}),  \notag \\
\mathbf{y}_{i} &=&\sqrt{\frac{m_{i}\left( m_{j}+m_{k}\right) }{%
(m_{i}+m_{j}+m_{k})\mu }}\left( \frac{m_{j}\mathbf{r}_{j}+m_{k}\mathbf{r}_{k}%
}{m_{j}+m_{k}}-\mathbf{r}_{i}\right) ,\text{ \ }i\neq j=1,2,3,
\label{Jacobi3}
\end{eqnarray}%
where

\begin{equation}
\mu =\sqrt{\frac{m_{i}m_{j}m_{k}}{m_{i}+m_{j}+m_{k}}}\ \   \label{EffecMass}
\end{equation}%
is the three-particle effective mass. 
In Eqs. (\ref{Jacobi3}) the subscripts $i$, $j$, and $k$ are a cyclic
permutation of the particle numbers. After the transformation (\ref{Jacobi3}%
) that allows the separation of c.m. and relative motions of three particles
with Hamiltonian (\ref{Trion}), the Schr\"{o}dinger equation for the
relative motion of the three-body system reads 
\begin{equation}
\left[ -\frac{\hbar ^{2}}{2\mu }(\nabla _{x_{i}}^{2}+\nabla
_{y_{i}}^{2})+V(x_{1})+V(x_{2})+V(x_{3})\right] \Psi (\mathbf{x}_{i},\mathbf{%
y}_{i})=E\Psi (\mathbf{x}_{i},\mathbf{y}_{i}).\ \   \label{Relative3}
\end{equation}%
In Eq. (\ref{Relative3}) $V(x_{i})$ is the interaction potential between two
particles 
at the relative distance $x_{1}$, $x_{2}$, and $x_{3}$, respectively, where $%
x_{i}$ is the modulus of the Jacobi vector $\mathbf{x}_{{i}}$ (\ref{Jacobi3}%
), and (\ref{Relative3}) is written for any of set $i=1,2,3$ of the Jacobi
coordinates (\ref{Jacobi3}). The orthogonal transformation between three
equivalent sets of the Jacobi coordinates (see, for example, \cite%
{FM,FilKez2024}) simplifies calculations of matrix elements involving $%
V(x_{i})$ potentials. The total wave function $\Psi (\mathbf{x}_{i},\mathbf{y%
}_{i})$ depends on one of the Jacobi coordinate sets $\mathbf{x}_{i},\mathbf{%
y}_{i}$. Generally, the trion wave function $\Psi (\mathbf{x}_{i},\mathbf{y}%
_{i})$ includes the electron-hole system spin function and the valence-band
wave function \cite{TMDCPRL2012}.

For the solution of the Schr\"{o}dinger equation (\ref{Relative3}) for the
trion in 2D configuration space, we use the method of hyperspherical
harmonics \cite{JibutiSh,Avery,Avery3,Avery2}. In the framework of this
method, the problem of the motion of three particles in 2D space interacting
via centrally symmetric potentials can be reduced to that of one particle in
4D space with an effective reduced mass (\ref{EffecMass}). Therefore,
instead of solving the three-body problem in 2D configuration space, we
solve a one-body problem in a four-dimensional hyperspace. Thus, for our
case, to solve Eq. (\ref{Relative3}) for a three-particle system in 2D
space, we turn to hyperspherical coordinates and hyperspherical harmonics,
which are the 4D generalizations of the familiar 3D spherical coordinates
and spherical harmonics. 
The main idea of the HH\ method is the solution of the Schr\"{o}dinger
equation (\ref{Relative3}) by separating the problem into two independent
parts: an angular part and a radial part. This can be done by the expansion
of the wave function of the trion $\Psi (\mathbf{x}_{i},\mathbf{y}_{i})$ in
terms of hyperspherical harmonics. The HH are the eigenfunctions of the
angular part of the generalized Laplace operator in the 4D configuration
space. The angular part of the problem is then solved analytically using
elegant theorems involving hyperspherical harmonics in 4D space \cite%
{Avery,Avery2}. For the radial part, we obtain the system of coupled
one-dimensional differential equations for the hyperradial functions\ that
can be solved numerically. In a diagonal approximation, for a particular
form of interparticle interactions, the system of coupled differential
equations can be solved analytically.

Due to the spherical symmetry\ of $V(x_{i})$ we employ hyperspherical
coordinates in 4D configuration space to obtain a solution of Eq. (\ref%
{Relative3}) for the trion. This allows us to separate the radial and
angular variables and reduce Eq. (\ref{Relative3}) for the system of coupled
one-dimensional radial equations. We introduce in 4D space the hyperradius $%
\rho =\sqrt{x_{i}^{2}+y_{i}^{2}}$ and a set of three angles $\Omega
_{i}\equiv (\alpha _{i},\varphi _{x_{i}},\varphi _{y_{i}}),$ where $\varphi
_{x_{i}}$ and $\varphi _{y_{i}}$ are the polar angles for 2D Jacobi vectors $%
\mathbf{x}_{i}$ and $\mathbf{y}_{j},$ respectively, and $\alpha _{i}$ is an
angle defined as $x_{i}=\rho \cos \alpha _{i},$ $y_{i}=\rho \sin \alpha
_{i}. $ After transforming from the Jacobi coordinates $\mathbf{x}_{i}$ and $%
\mathbf{y}_{i}$ to the hyperspherical coordinates $\rho $ and $\Omega _{i}$,
and using the familiar expression for the Laplacian operator in the 4D
configuration space \cite{Avery,Avery2}, Eq. (\ref{Relative3}) can be
rewritten as

\begin{equation}
\left[ -\frac{\hbar ^{2}}{2\mu }\left( \frac{\partial ^{2}}{\partial
^{2}\rho }+\frac{3}{\rho }\frac{\partial }{\partial \rho }-\frac{\widehat{K}%
^{2}(\Omega _{i})}{\rho ^{2}}\right) +\sum_{j=1}^{3}V(\rho \cos \alpha
_{j})-E\right] \Psi (\rho ,\Omega _{i})=0,\ \   \label{HHtrion}
\end{equation}%
where $\widehat{K}^{2}(\Omega _{i})$ is the angular part of the generalized
Laplace operator in 4D configuration space known as the grand-angular
momentum operator \cite{JibutiSh,Avery,Avery2}

\begin{eqnarray}
\widehat{K}^{2}(\Omega _{i}) &=&\frac{d^{2}}{d\alpha _{i}^{2}}+2\cot 2\alpha
_{i}\frac{d}{d\alpha _{i}}-\frac{\widehat{l}^{2}(\varphi _{x_{i}})}{\cos
\alpha _{i}}-\frac{\widehat{l}^{2}(\varphi _{y_{i}})}{\sin \alpha _{i}}, 
\notag \\
\widehat{l}(\varphi _{x_{i}})\ \ &=&-i\frac{d}{d\varphi _{x_{i}}},\text{ \ }%
\widehat{l}(\varphi _{y_{i}})\ \ =-i\frac{d}{d\varphi _{y_{i}}},\text{ }%
\Omega _{i}\equiv (\alpha _{i},\widehat{\mathbf{x}}_{i},\widehat{\mathbf{y}}%
_{i}).
\end{eqnarray}%
We shall seek the common eigenfunctions of the operator $\widehat{K}%
^{2}(\Omega _{i})$ and (\ref{HHtrion}). One can expand the wave function of
the trion $\Psi (\rho ,\Omega _{i})$ in terms of the HH that are the
eigenfunctions of the operator $\widehat{K}^{2}$: 
\begin{equation}
\widehat{K}^{2}(\Omega )\Phi _{K}^{l_{x}m_{x}l_{y}m_{y}}(\Omega )=K(K+2)\Phi
_{K}^{l_{x}m_{x}l_{y}m_{y}}(\Omega ).\ 
\end{equation}%
The grand-angular momentum $K=2n+l_{x}+l_{y}$, where $l_{x}$ and $l_{y}$ are
angular momenta corresponding to the Jacobi coordinates $\mathbf{x}$ and $%
\mathbf{y,}$ respectively, and $n$ $\geqslant 0$ is an integer number. The
eigenfunctions $\Phi _{K}^{l_{x}m_{x}l_{y}m_{y}}(\Omega )$ present a
complete set of the orthonormal basis. The function $\Phi
_{K}^{l_{x}l_{y}LM}(\Omega )\equiv $ $\Phi _{K\lambda }^{L}(\Omega )$ (here
and below \ we use the short-hand notation $\lambda \equiv $ $%
\{l_{x},l_{y}\} $) with the total orbital angular momentum of the trion $L$
and its projection $M$, is a linear combination of HH $\Phi
_{K}^{l_{x}m_{x}l_{y}m_{y}}(\Omega )$, given by:

\begin{equation}
\Phi _{K\lambda }^{L}(\Omega )=\sum_{m_{x}m_{y}}\left\langle
l_{x}m_{x}l_{y}m_{y}\right. \left\vert LM\right\rangle \Phi
_{K}^{l_{x}m_{x}l_{y}m_{y}}(\Omega ),\   \label{Clebsh}
\end{equation}%
where $\left\langle l_{x}m_{x}l_{y}m_{y}\right. \left\vert LM\right\rangle $
are the Clebsch--Gordan coefficients. The equation (\ref{Clebsh}) is valid
for each partition $i$ for the Jacobi coordinates. The $\Phi _{K\lambda
}^{L}(\Omega )$\ are also the eigenfunctions of the operator $\widehat{K}%
^{2} $. The expansion of $\Psi (\rho ,\Omega _{i})$ in terms of $\Phi
_{K\lambda }^{L}(\Omega )$ reads

\begin{equation}
\Psi (\rho ,\Omega _{i})=\rho ^{-3/2}\sum_{_{K\lambda }}u_{K\lambda
}^{L}(\rho )\Phi _{K\lambda }^{L}(\Omega ).\   \label{ExpanTrion}
\end{equation}%
In Eq. (\ref{ExpanTrion}) $u_{K\lambda }^{L}(\rho )$ are the hyperradial
functions for the trion state with the total orbital angular momentum $L$. 
By substituting (\ref{ExpanTrion}) into (\ref{HHtrion}), then multiplying on
the left by $\Phi _{K\lambda }^{L\ast }(\Omega _{i})$ and integrating over
the hyperangles, we separate the radial and angular variables and obtain the
set of coupled differential equations for the hyperradial functions $%
u_{K\lambda }^{L}(\rho )$: 
\begin{equation}
\left[ \frac{d^{2}}{d\rho ^{2}}-\frac{(K+1)^{2}-1/4}{\rho ^{2}}+\kappa ^{2}%
\right] u_{K\lambda }^{L}(\rho )=\frac{2\mu }{\hbar ^{2}}\sum_{_{K^{^{\prime
}}\lambda ^{^{\prime }}}}\mathcal{W}_{K\lambda K^{^{\prime }}\lambda
^{^{\prime }}}^{L}(\rho )u_{K^{^{\prime }}\lambda ^{^{\prime }}}^{L}(\rho ).
\label{TrionGeneral}
\end{equation}%
In Eq. (\ref{TrionGeneral}) $\kappa ^{2}=2\mu B/\hslash ^{2},$ where $B$ is
the trion binding energy, and the effective potential energy $\mathcal{W}%
_{K\lambda K^{^{\prime }}\lambda ^{^{\prime }}}^{L}(\rho )$ is 
\begin{equation}
\mathcal{W}_{K\lambda K^{^{\prime }}\lambda ^{^{\prime }}}^{L}(\rho )=\int
\Phi _{K\lambda }^{L\ast }(\Omega _{1})\sum_{j=1}^{3}V(\rho \cos \alpha
_{j})\Phi _{K^{^{\prime }}\lambda ^{^{\prime }}}^{L}(\Omega _{1})d\Omega
_{1}.  \label{W3general}
\end{equation}%
The coupled system of equations (\ref{TrionGeneral}) describes the motion of
one particle in the centrally symmetric coupling effective field (\ref%
{W3general}). The coupling effective interaction (\ref{W3general}) is
defined via the RK potential~(\ref{Keldysh}). Substituting (\ref{Keldysh})
into Eq. (\ref{W3general}), one obtains 
\begin{eqnarray}
\mathcal{W}_{K\lambda K\lambda ^{^{\prime }}}^{L}(\rho ) &=&\frac{\pi ke^{2}%
}{2\epsilon \rho _{0}}\left\{ \int \Phi _{K\lambda }^{L\ast }(\Omega _{1})%
\left[ H_{0}(\frac{x_{1}}{b_{j}\rho _{0}})-Y_{0}(\frac{x_{1}}{b_{j}\rho _{0}}%
)\right] \Phi _{K^{^{\prime }}\lambda ^{^{\prime }}}^{L}(\Omega _{1})d\Omega
_{1}\right. -  \notag \\
&&\int \Phi _{K\lambda }^{L\ast }(\Omega _{1})\left[ H_{0}(\frac{x_{2}}{%
b_{j}\rho _{0}})-Y_{0}(\frac{x_{2}}{b_{j}\rho _{0}})\right] \Phi
_{K^{^{\prime }}\lambda ^{^{\prime }}}^{L}(\Omega _{1})d\Omega _{1}-  \notag
\\
&&\left. \int \Phi _{K\lambda }^{L\ast }(\Omega _{1})\left[ H_{0}(\frac{x_{3}%
}{b_{j}\rho _{0}})-Y_{0}(\frac{x_{3}}{b_{j}\rho _{0}})\right] \Phi
_{K^{^{\prime }}\lambda ^{^{\prime }}}^{L}(\Omega _{1})d\Omega _{1}\right\} .
\label{Keldysh W}
\end{eqnarray}%
where 
\begin{eqnarray}
b_{1} &=&\sqrt{\frac{m_{2}m_{3}}{(m_{2}+m_{3})\mu }\ },\text{ }\ b_{2}=\sqrt{%
\frac{m_{1}m_{3}}{(m_{1}+m_{3})\mu }\ },\text{ \ }b_{3}=\sqrt{\frac{%
m_{1}m_{2}}{(m_{1}+m_{2})\mu }\ },  \label{b1b2b3} \\
x_{1} &=&\rho \cos \alpha _{1},\text{ \ }x_{2}=\rho \cos \alpha _{2},\text{
\ }x_{3}=\rho \cos \alpha _{3};\text{ \ }d\Omega _{i}=\cos \alpha _{i}\sin
\alpha _{i}\text{ }d\alpha _{i}d\varphi _{x_{i}}d\varphi _{y_{i}}.
\label{x1x2x3}
\end{eqnarray}%
The $X^{-}$ and $X^{+}$ trions have two identical particles, two electrons
and two holes, respectively. Below we always assume that indices 2 and 3
identify the identical particles. The total wave function of the system \ is
antisymmetric with respect to the permutation of the identical particles.
Calculation of matrix elements $\mathcal{W}_{K\lambda K\lambda ^{^{\prime
}}}(\rho )$ of the two-body $V(\rho \cos \alpha _{j})$ interactions in the
hyperspherical harmonics expansion method for a three-body system is greatly
simplified by using the HH basis states appropriate for the partition
corresponding to the interacting pair. There are three terms in the integral
(\ref{Keldysh W}). From these three terms, it is relatively easy to
calculate only one term whose index of the $x_{i}$ coincides with the index
of the set of angles on which the hyperspherical functions $\Phi
_{K^{^{\prime }}\lambda ^{^{\prime }}}^{L}(\Omega _{i})$ depend. In this
case the evaluation of the integrals by the polar angles for the Jacobi
vectors $\mathbf{x}_{i}$ and $\mathbf{y}_{i}$ can be done analytically in a
complete form using the properties of hyperspherical harmonics. Direct
calculations of the other two terms are a challenging task. To simplify the
calculations, it is convenient to use a unitary transformation that involves
the Reynal-Revai coefficients (RRC) \cite{Revai}, which are the
transformation coefficients between the HH bases corresponding to the two
partitions and give the relationship between the HH defined on the two
different sets of Jacobi coordinates. The use of RRC becomes particularly
essential for the numerical solution of the three-body Schr\H{o}dinger
equation where the two-body potentials do not allow analytical integration
by the angles. The Reynal-Revai unitary transformation between different
bases of HH functions (transformation from the HH basis for the partition $k$
to the HH basis for the partition $i$) has the following form 
\begin{equation}
\Phi _{K\lambda _{i}}^{L}(\Omega _{i})=\sum\limits_{\lambda
_{k}}\left\langle \lambda _{k}\right\vert \lambda _{i}\rangle _{KL}\Phi
_{K\lambda _{k}}^{L}(\Omega _{k}),  \label{RRC}
\end{equation}%
where $\left\langle \lambda _{k}\right\vert \lambda _{i}\rangle _{KL}\equiv $
$\left\langle l_{x_{k}}l_{y_{k}}\right\vert l_{x_{i}}l_{y_{i}}\rangle _{KL}$
are Reynal-Revai coefficients. The Reynal-Revai unitary transformation does
not change the global momentum $K$ and the total angular momentum $L$ of the
three-particle system. By using the unitary transformation (\ref{RRC}), Eq. (%
\ref{Keldysh W}) can be written in the following form 
\begin{equation}
W_{KK^{^{\prime }}\text{ }\lambda \lambda ^{\prime }}^{L}(\rho )=\left( 
\mathcal{K}_{K\lambda K^{^{\prime }}\lambda ^{^{\prime
}}}^{(1)}-\sum\limits_{\lambda _{k}\lambda _{k}^{^{\prime }}}\left\langle
\lambda _{k}\right\vert \lambda _{i}\rangle _{KL}\left\langle \lambda
_{k}^{^{\prime }}\right\vert \lambda _{i}\rangle _{K^{\prime }L}\mathcal{K}%
_{K\lambda K^{^{\prime }}\lambda ^{^{\prime }}}^{(2)}-\sum\limits_{\lambda
_{j}\lambda _{j}^{^{\prime }}}\left\langle \lambda _{j}\right\vert \lambda
_{i}\rangle _{KL}\left\langle \lambda _{j}^{^{\prime }}\right\vert \lambda
_{i}\rangle _{K^{\prime }L}\mathcal{K}_{K\lambda K^{^{\prime }}\lambda
^{^{\prime }}}^{(3)}\right) ,  \label{WKeldysh}
\end{equation}%
where 
\begin{equation}
\mathcal{K}_{K\lambda K^{^{\prime }}\lambda ^{^{\prime }}}^{(i)}=\frac{\pi
ke^{2}}{2\epsilon \rho _{0}}\int \Phi _{K\lambda }^{L\ast }(\Omega _{i})%
\left[ H_{0}(\frac{\rho \cos \alpha _{i}}{b_{i}\rho _{0}})-Y_{0}(\frac{\rho
\cos \alpha _{i}}{b_{i}\rho _{0}})\right] \Phi _{K^{^{\prime }}\lambda
^{^{\prime }}}^{L}(\Omega _{i})\cos \alpha _{i}\sin \alpha _{i}\text{ }%
d\alpha _{i}d\varphi _{x_{i}}d\varphi _{y_{i}},\text{ }i=1,2,3.
\label{K_avarage}
\end{equation}%
Using the matrix elements of the effective potential energies (\ref%
{K_avarage}), one can solve the coupled differential equations (\ref%
{TrionGeneral}) numerically. However, one can obtain analytical solutions
for the trion in the diagonal approximation for a long- and short-range
approximations of the 2D screened charge-charge interactions given by Eq. (%
\ref{Limit}). The results of the approximate analytical and numerical
solutions of Eq. (\ref{TrionGeneral}) are presented in the next sections.

\section{Approximate analytical solutions}

The finding of the binding energy and wave function for a trion requires the
solution of the system of equations (\ref{TrionGeneral}) coupled via the
potential energy (\ref{WKeldysh}). Many calculations in atomic and nuclear
physics that have been made within the framework of the HH method use
approximations for the solution of the coupled differential equations for
the hyperradial functions. In this Section, we obtain analytical solutions
for the trion in the diagonal approximation for both long- and short-range
approximations of the 2D screened charge-charge interaction. In the diagonal
approximation, the coupled differential equations (\ref{TrionGeneral}) will
be decoupled.

Let us consider the solution of Eq. (\ref{TrionGeneral}) with the effective
potential energy (\ref{WKeldysh}) for the long- and short-range interactions
(\ref{Limit}). The matrix elements (\ref{WKeldysh}) for a central potential
when $K=K^{^{\prime }}$ and $\lambda =\lambda ^{^{\prime }} $ (diagonal
matrix elements) are much larger than non-diagonal matrix elements and make
the main contribution to the binding energy. From the other side, if one
neglects the non-diagonal matrix elements, the differential equations (\ref%
{TrionGeneral}) will be decoupled. It is easier to solve decoupled equations
numerically with much less computational complexity. Moreover, for some
central potentials, the diagonal approximation even allows to obtain an
analytical solution for the decoupled equations. Below we obtain analytical
solutions of Eq. (\ref{TrionGeneral}) in the limits $r>>\rho _{0}$ and $%
r<<\rho _{0}$ for the potentials (\ref{Limit}) in the diagonal approximation.

\subsection{$\protect\bigskip $Long-range limit:$\ r>>\protect\rho _{0}$}

Consider the long range $r>>\rho _{0}$ limit when the Keldysh potential can
be approximated by the Coulomb potential (\ref{Limit}). In this case, the
effective potential energy ~(\ref{W3general}) can be written as

\begin{equation}
\mathcal{W}_{K\lambda K^{^{\prime }}\lambda ^{^{\prime }}}^{L}(\rho )=k\int
\Phi _{K\lambda }^{L\ast }(\Omega _{1})\left( \frac{b_{1}}{x_{1}}-\frac{b_{2}%
}{x_{2}}-\frac{b_{3}}{x_{3}}\right) \Phi _{K^{^{\prime }}\lambda ^{^{\prime
}}}^{L}(\Omega _{1})d\Omega _{1}.  \label{CoulombPotential}
\end{equation}%
In Eq. (\ref{CoulombPotential}) $b_{1},$ $b_{2},$ $b_{3}$ and $x_{1},$ $%
x_{2},$ $x_{3}$ are defined in Eqs. (\ref{b1b2b3}) and (\ref{x1x2x3}),
respectively. By using the Reynal-Revai unitary transformation (\ref{RRC})
for the fixed global momentum $K$ and the total angular momentum $L$, as
well as Eq. (\ref{x1x2x3}), one can rewrite (\ref{CoulombPotential}) in the
following form

\begin{equation}
\mathcal{W}_{K\lambda K^{^{\prime }}\lambda ^{^{\prime }}}^{L}(\rho )=\frac{%
\mathcal{G}_{K\lambda K^{^{\prime }}\lambda ^{^{\prime }}}^{L}}{\rho },
\end{equation}%
where %
\begin{equation}
\mathcal{G}_{K\lambda K^{^{\prime }}\lambda ^{^{\prime }}}^{L}=ke^{2}\left(
b_{1}\mathcal{C}_{K\lambda K^{^{\prime }}\lambda ^{^{\prime
}}}^{(1)}-b_{2}\sum\limits_{\lambda _{k}\lambda _{k}^{^{\prime
}}}\left\langle \lambda _{k}\right\vert \lambda _{i}\rangle
_{KL}\left\langle \lambda _{k}^{^{\prime }}\right\vert \lambda _{i}\rangle
_{K^{\prime }L}\mathcal{C}_{K\lambda K^{^{\prime }}\lambda ^{^{\prime
}}}^{(2)}-b_{3}\sum\limits_{\lambda _{j}\lambda _{j}^{^{\prime
}}}\left\langle \lambda _{j}\right\vert \lambda _{i}\rangle
_{KL}\left\langle \lambda _{j}^{^{\prime }}\right\vert \lambda _{i}\rangle
_{K^{\prime }L}\mathcal{C}_{K\lambda K^{^{\prime }}\lambda ^{^{\prime
}}}^{(3)}\right) ,  \label{gCoulomb}
\end{equation}

and

\begin{equation}
\mathcal{C}_{K\lambda K^{^{\prime }}\lambda ^{^{\prime }}}^{(i)}=\int \Phi
_{K\lambda }^{L\ast }(\Omega _{i})\left( \frac{1}{\cos \alpha _{i}}\right)
\Phi _{K^{^{\prime }}\lambda ^{^{\prime }}}^{L}(\Omega _{i})d\Omega _{i},%
\text{ }i=1,2,3.  \label{CoulombIntergral}
\end{equation}%
All integrals (\ref{CoulombIntergral}) can be evaluated in a closed analytic
form using the properties of HH \cite{JibutiSh,Avery,Avery2} and the
corresponding expressions are presented in Appendix A.

Consider the diagonal approximation when $K=K^{^{\prime }},$ $\lambda
=\lambda ^{^{\prime }}.$ The latter leads to the reduction of the system of
equations (\ref{TrionGeneral}) to a simpler form. In this case, the system
of equations (\ref{TrionGeneral}) will be decoupled and reduced to the
following differential equation:

\begin{equation}
\left[ \frac{d^{2}}{d\rho ^{2}}-\frac{(K+1)^{2}-1/4}{\rho ^{2}}+\kappa ^{2}%
\right] u_{K\lambda }^{L}(\rho )=\frac{2\mu }{\hbar ^{2}}\frac{\mathcal{G}%
_{K\lambda K\lambda }^{L}}{\rho }u_{K\lambda }^{L}(\rho ).
\label{2D Coulomb}
\end{equation}%
Eq. (\ref{2D Coulomb}) has the same dependence on the variable $\rho $ as
the 3D radial Schr\"{o}dinger equation for a hydrogen atom. Following the
standard procedure for the solution of the radial Schr\"{o}dinger equation
for a hydrogen atom in 3D space \cite{LandauQM}, which was generalized for
2D space \cite{Zaslow1967,Yang1991,Portnoi2002,KezFBS2023}, one can solve
Eq. (\ref{2D Coulomb}) analytically and obtain eigenenergies and
eigenfunctions

\begin{equation}
E=-\frac{\mu }{\hbar ^{2}}\frac{\left( \mathcal{G}_{K\lambda K\lambda
}^{L}\right) ^{2}}{2(N+K+3/2)^{2}},\ \ \   \label{Colim}
\end{equation}

\ \ 
\begin{equation}
u_{K}(\rho )=C_{K}\rho ^{K+3/2}\exp (-\kappa \rho )L_{N}^{2K+2}(2\kappa \rho
)
\end{equation}%
where $N=1,2,...$ , $L_{N}^{2K+2}$ is a Laguerre polynomial, $C_{K}$ is a
normalization constant and $\mathcal{G}_{K\lambda K\lambda }^{L}$ are
defined by (\ref{gCoulomb}) when $K=K^{^{\prime }},$ $\lambda =\lambda
^{^{\prime }}$. In this case (\ref{gCoulomb}) becomes much more compact.

\subsection{\protect\bigskip Short-range limit: $r<<\protect\rho _{0}$}

Now let us consider a short-range limit $r<<\rho _{0}$ in which the RK
potential can be approximated by the logarithmic potential (\ref{Limit}).
The logarithmic potential is very widely used as a quark confinement
potential in high energy physics. The solution of the Schr\"{o}dinger
equation with the logarithmic potential is a very complex problem. There are
only almost complete solutions of the Schr\"{o}dinger equation for the
logarithmic potential \cite{Gesztesy1978,Muller1979,Khelashvili1}. In Ref. 
\cite{Ganchev} a remarkable solution for the trion with a logarithmic
potential was obtained, but only for the case, where the full Rytova-Keldysh
effective potential is replaced with a completely logarithmic form. 
Below, we consider the logarithmic potential, which has a shift due to the
asymptotic approximation of the RK potential at $r<<\rho _{0}$, as it is
seen from Eq. (\ref{Limit}). Using the corresponding Jacobi coordinates, the
effective potential energy (\ref{W3general}) can be written as

\begin{equation}
\mathcal{W}_{K\lambda K^{^{\prime }}\lambda ^{^{\prime }}}(\rho )=\frac{%
ke^{2}}{\epsilon \rho _{0\text{ }}}\int \Phi _{K\lambda }^{L\ast }(\Omega
_{1})\left( (\ln \frac{x_{1}}{2b_{1}\rho _{0\text{ }}}+\gamma )-(\ln \frac{%
x_{2}}{2b_{2}\rho _{0\text{ }}}+\gamma )-(\ln \frac{x_{3}}{2b_{3}\rho _{0%
\text{ }}}+\gamma )\right) \Phi _{K^{^{\prime }}\lambda ^{^{\prime
}}}^{L}(\Omega _{1})d\Omega _{1},
\end{equation}%
where $b_{1},$ $b_{2},$ $b_{3},$ and $x_{1},$ $x_{2}$, $x_{3}$ are defined
in Eqs. (\ref{b1b2b3}) and (\ref{x1x2x3}), respectively. The effective
potential energy can be evaluated by employing the Reynal-Revai unitary
transformation. As a result, we have

\begin{equation}
\mathcal{W}_{K\lambda K^{^{\prime }}\lambda ^{^{\prime }}}^{L}(\rho )=-\frac{%
ke^{2}}{\epsilon \rho _{0\text{ }}}\ln \frac{\rho }{\rho _{0\text{ }}}+%
\mathcal{B}_{123}+\mathcal{J}_{K\lambda K^{^{\prime }}\lambda ^{^{\prime
}}}^{L}\text{,}  \label{W Log}
\end{equation}%
where

\begin{eqnarray}
\mathcal{B}_{123} &=&\frac{ke^{2}}{\epsilon \rho _{0\text{ }}}(\ln \frac{%
2b_{2}b_{3}}{b_{1}}-\gamma ),  \label{B123} \\
\mathcal{J}_{K\lambda K^{^{\prime }}\lambda ^{^{\prime }}}^{L} &=&\frac{%
ke^{2}}{\epsilon \rho _{0\text{ }}}(\mathcal{L}_{K\lambda K^{^{\prime
}}\lambda ^{^{\prime }}}^{(1)}-\sum\limits_{\lambda _{k}\lambda
_{k}^{^{\prime }}}\left\langle \lambda _{k}\right\vert \lambda _{i}\rangle
_{KL}\left\langle \lambda _{k}^{^{\prime }}\right\vert \lambda _{i}\rangle
_{KL}\text{ }\mathcal{L}_{K\lambda K^{^{\prime }}\lambda ^{^{\prime
}}}^{(2)}-\sum\limits_{\lambda _{j}\lambda _{j}^{^{\prime }}}\left\langle
\lambda _{j}\right\vert \lambda _{i}\rangle _{KL}\left\langle \lambda
_{j}^{^{\prime }}\right\vert \lambda _{i}\rangle _{KL}\mathcal{L}_{K\lambda
K^{^{\prime }}\lambda ^{^{\prime }}}^{(3)}).  \label{J Log}
\end{eqnarray}%
In Eq. (\ref{J Log}) $\mathcal{L}_{K\lambda K^{^{\prime }}\lambda ^{^{\prime
}}}^{(i)}$ is defined as

\begin{equation}
\mathcal{L}_{K\lambda K^{^{\prime }}\lambda ^{^{\prime }}}^{(i)}=\int \Phi
_{K\lambda }^{L\ast }(\Omega _{i})\ln \cos \alpha _{i}\Phi _{K^{^{\prime
}}\lambda ^{^{\prime }}}^{L}(\Omega _{i})\text{{}}d\Omega _{i},\text{ \ }%
i=1,2,3.  \label{LKK}
\end{equation}%
and can be found in a closed analytic form for any $K\lambda K^{^{\prime
}}\lambda ^{^{\prime }}$ set and corresponding expression for (\ref{LKK}) is
presented in Appendix B.

Substituting (\ref{W Log}) into Eq. (\ref{TrionGeneral}), one obtains the
coupled differential equations for the hyperradial functions $u_{K\lambda
}(\rho )$ for the trion in the asymptotic region $r<<\rho _{0}$. The latter
system of equations can be written in a diagonal approximation when $%
K=K^{^{\prime }}$ and $\lambda =\lambda ^{^{\prime }}$, leading to the
following equation

\begin{equation}
\left[ \frac{d^{2}}{d\rho ^{2}}-\frac{(K+1)^{2}-1/4}{\rho ^{2}}+\kappa ^{2}%
\right] u_{K\lambda }^{L}(\rho )=\frac{2\mu }{\hbar ^{2}}\left( -\frac{ke^{2}%
}{\epsilon \rho _{0\text{ }}}\ln \frac{\rho }{\rho _{0\text{ }}}+\mathcal{B}%
_{123}+\mathcal{J}_{K\lambda K\lambda }^{L}\right) u_{K\lambda }^{L}(\rho ).
\label{LogDiagonal}
\end{equation}%
Equation (\ref{LogDiagonal}) can be rewritten in the following form

\begin{equation}
\left[ \frac{d^{2}}{d\rho ^{2}}+\left( \alpha -\beta \text{ln}\rho -\frac{%
(K+1)^{2}-1/4}{\rho ^{2}}\right) \right] u_{K\lambda }^{L}(\rho )=0.
\label{RewEq}
\end{equation}%
Here we have defined

\begin{equation}
\alpha =\kappa ^{2}(2\mu B/\hslash ^{2})+\frac{2\mu }{\hbar ^{2}}\left( 
\frac{ke^{2}}{\epsilon \rho _{0\text{ }}}\ln \rho _{0\text{ }}-\mathcal{B}%
_{123}-\mathcal{J}_{K\lambda K\lambda }\right) ,\text{ \ \ \ }\beta =\frac{%
2\mu }{\hbar ^{2}}\frac{ke^{2}}{\epsilon \rho _{0\text{ }}}.
\end{equation}%
Note that $\mathcal{J}_{K\lambda K\lambda }^{L}$ is defined in Eq. (\ref{W
Log}) when $K=K^{^{\prime }},$ $\lambda =\lambda ^{^{\prime }}$. In this
case Eq. (\ref{W Log}) becomes much more compact.

One can find the approximate analytical solution of Eq. (\ref{RewEq}) by
rescaling the variable $\rho $ and introducing a new function that allows us
to reduce Eq. (\ref{RewEq}) to the known Weber's equation \cite{Weber 1869}.
The reduction of Eq. (\ref{RewEq}) to the Weber's equation is given in
Appendix C. The solution of this equation in a first approximation is simply
the parabolic cylinder functions $D_{n}.$ The eigenvalues that correspond to
these eigenfunctions are

\begin{equation}
E=-\frac{ke^{2}}{\epsilon \rho _{0\text{ }}}\ln \rho _{0\text{ }}+\mathcal{B}%
_{123}+\mathcal{J}_{K\lambda K\lambda }+\frac{ke^{2}}{\epsilon \rho _{0\text{
}}}\left\{ 1+\ln \left[ \frac{[n+1/2+\sqrt{(n+1/2)^{2}+(K+2)^{2}/16}]^{2}}{%
\beta }\right] \right\} ,\text{ }n=0,1,2...  \label{Kelim}
\end{equation}

\section{Application to TMDCs: Results and discussion}

\begin{figure}[h]
\centering
\includegraphics[width=11.0cm]{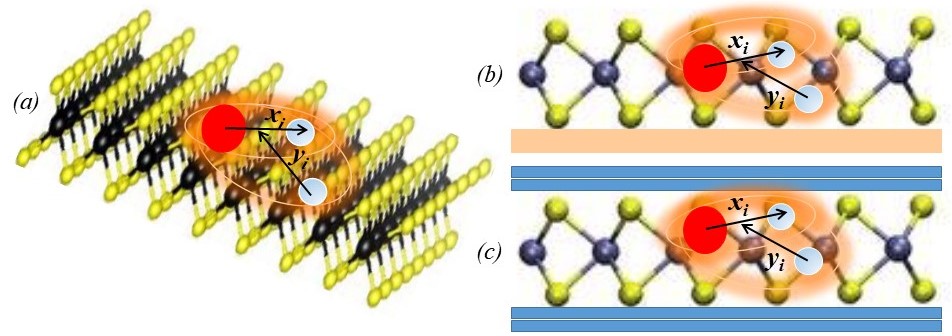}
\caption{(Color online) Schematic representation of the trion in ($a$)
freestanding, ($b$) supported and ($c$) encapsulated TMDC monolayer. $x_{i}$
and $y_{i}$ are Jacobi coordinates for the partition $i$. }
\label{TrionMon}
\end{figure}

We apply the present theoretical approach for calculations of the trion
binding energies in the following TMDC monolayers:\ MoS$_{2}$, MoSe$_{2}$, WS%
$_{2}$, and WSe$_{2}$. A schematic representation of a trion in
freestanding, supported, and encapsulated TMDC monolayers is given in Fig. %
\ref{TrionMon}. The form of the trion wave function (\ref{ExpanTrion}) is
most general, not restricted to any particular mass ratio of electrons and
holes, and describes the three-particle relative motion. However, two from
three particles constituting a positive or a negatively trion in the
monolayer plane must be identical, as shown in Fig. \ref{TrionMon}. Two
identical particles must have different spin and/or valley indices and a
trion is formed by the singlet or triplet excitons. Therefore, it can be in
either a single or triplet state. 

\subsection{Intravalley and intervalley trions}

Trions play an important role in the fundamental valley dynamics in emerging
two-dimensional semiconductor materials. Trions properties in 2D TMDC are
associated with their spin and valley degrees of freedom. The lack of
inversion symmetry, together with a strong spin-orbit coupling, leads to a
large valence band splitting in monolayer TMDCs \cite{TMDCPRL2012}. This
splitting lifts the spin degeneracy of both electron and hole states in the $%
K$ and $K^{\prime }$ valleys, enabling the excitation of carriers with
various combinations of spin and valley indices. The valley degree of
freedom has been vastly explored in monolayers of molybdenum- and
tungsten-based dichalcogenides using helicity-resolved photoluminescence,
time-resolved photoluminescence, reflectance spectra and the time-resolved
Kerr rotation technique \cite{Mak2012,Cao2012,WSe2
Jones,Zeng2012,Sallen2012,bZu,Wang2015,MacNeill2015,Gao2016,Yan2016,Huang2017,Hao2017}%
. The absolute minima of the conduction and absolute maxima of the valence
bands in TMDC monolayers are at two non-equivalent hexagonal Brillouin zone
corners of the $K$ and $K^{\prime }$ valleys. The inversion symmetry
breaking with the spin-orbit interaction leads to the carrier spin and
valley coupling so that the circular polarization of the absorbed or emitted
photon can be directly associated with a specific valley, $K$ or $K^{\prime
} $ \cite{TMDCPRL2012,Sallen2012}. This valley contrasting optical selection
rules allow one to address and manipulate the valley index. Experimental
studies \cite{Mak2012,Cao2012} show that each of the valleys can be excited
by radiation of given helicity. Having a valley index $\tau $ in addition to
spin, one can distinguish between two types of trions: the intravalley and
intervalley trions. Our theoretical approach allows consideration of this
degree of freedom. For intravalley trions $\tau _{1}=\tau _{2}$, while for
intervalley trions $\tau _{1}\neq \tau _{2}$. A symmetry analysis of the
different $X^{-}$ and $X^{+}$ trion states, classified by the valley
configurations as well as the spin configuration, is performed \cite%
{Ganchev,CourtadeSemina,NScRev2015,Fu2019}. Depending on the spin
configuration of the electron--hole pairs, intravalley excitons of TMDC
monolayers can be either optically bright or dark. The analysis \cite%
{CourtadeSemina,NScRev2015,Hao2017,Fu2019} shows that the negative and
positive trions can be optically bright or dark. Experimental studies \cite%
{DarkT2017,Liu2019} have revealed evidence of dark trions in monolayer WSe$%
_{2}$ under a strong in-plane magnetic field.

\begin{figure}[h]
\caption{(Color online) Schematic illustration of WSe$_{2}$ low-energy band
structure and the spin-valley configurations of the constituent charge
carriers. The topmost spin-subband for the valence band and the lower and
upper spin-orbit splitting conduction band are shown. Arrows denote bands
with up (down) spin. The hole has the opposite spin of the valence electron 
\protect\cite{Robert2017}. Light and dark rectangles indicate the bright and
dark trions, respectively. $(a)$, $(b)$, $(c)$, $(d)$, and $(e)$ correspond
to $X^{-}$ trions. $(g)$ and $(h)$ correspond to $X^{+}$ trions. The bright
trions emit circularly polarized light in the out-of-monolayer plane
direction, while the dark trions emit vertically polarized light in the
in-monolayer plane direction \protect\cite{Liu2019}. Lines indicate
interaction between three charged particles. The remaining configurations
are the time reversal of those shown in the figure.}
\label{TrionsValle}\centering
\includegraphics[width=13.0cm]{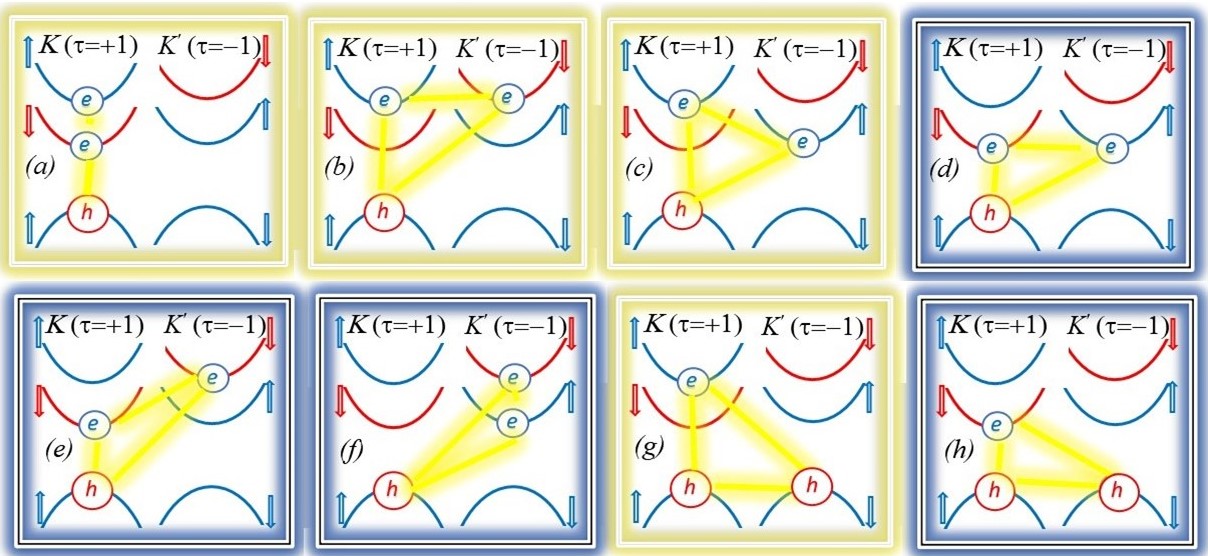}
\par
\end{figure}

The atomic spin-orbit interaction in TMDC introduces large spin-orbit
splitting of the valence band, so that the low-energy trions involve only
the lowest hole states \cite{Mosttop,Mosttop2}, because in the $K$ and $%
K^{\prime }$ valleys, the spin--orbit splitting of the valence band is much
larger than that of the conduction band \cite{TMDCPRL2012}. The developed
theoretical formalism allows us to consider a three-band model, consisting
of a higher spin branch of the valence band and both spin branches of the
conduction band. We consider only one possible hole spin state at the
topmost valence band in a given valley \cite{Mosttop,Mosttop2}: spin up in
the $K$ valley and spin down in the $K^{\prime }$ valley. However, because
the spin-orbit interaction for electrons is much smaller than the trion
binding energy, one should take for each electron four states coming from
the spin and valley degeneracy. The negative trions can be bright or dark.
When the electron and hole come from bands with the same electron spin,
their recombination emits light. This is a bright exciton. When the electron
is in the bottom spin-orbit split state, the spin of the electron and hole
mismatch. The spin mismatch strongly suppresses the radiative recombination
of the electron and hole. This is a dark exciton. In finite charge density,
bright and dark excitons can interact with the Fermi sea to form bright and
dark trions. Dark trions were predicted in Refs. \cite%
{Deilmann2017,Danovich2017}. The intravalley\ $X^{-}$ trion in the
spin-singlet state, with both electrons and a hole in the same valley, is a
bright spin singlet trion (Fig. \ref{TrionsValle}$a$). The intervalley
trions with electrons in two different valleys can be bright (Fig. \ref%
{TrionsValle}$b$ and \ref{TrionsValle}$c$) or dark (Fig. \ref{TrionsValle}$%
d,e,f$), depending on the spin of the electrons state in the different
valleys. Thus, the intravalley $X^{-}$ trion is in a spin singlet state,
while the intervalley $X^{-}$ trion can be in a spin singlet or spin triplet
state. Consequently, $X^{-}$ can be of both spin-forbidden and
momentum-forbidden types. The most intriguing feature is that in
cross-circularly polarized experiments, trions created in the $K$ valley in
the singlet state (Fig. \ref{TrionsValle}$a$) are converted to intervalley
singlet trions with an electron in the $K^{\prime }$ valley (Fig. \ref%
{TrionsValle}$c$) via the spin flip and electron-hole exchange interaction\ 
\cite{Singh2016}.

The positive trion is formed of two holes occupying the topmost valence band
sub bands and the electron in conduction band sub bands is always the
intervalley trion. $X^{+}$ is a bright trion when the spin orientation of
the electron and hole is the same (Fig. \ref{TrionsValle}$g$). Otherwise, $%
X^{+}$ is dark trion (Fig. \ref{TrionsValle}$h$).

The $X^{-}$ trion consists of two electrons of opposite spin in the two
conduction bands as well as one hole in the topmost valence band.
Consequently, the intravalley $X^{-}$ trion ($\tau _{1}=\tau _{2}$) must
exist in the spin-singlet state, whereas the intervalley $X^{-}$ trion ($%
\tau _{1}\neq \tau _{2}$) can exist in either the singlet or triplet states.
Intravalley $X^{-}$ trions have been shown to be optically bright
(radiative), provided that the spins of a paired electron and hole differ.
In contrast, intervalley $X^{-}$ trions in TMDC monolayers such as the
molybdenum- and tungsten-based dichalcogenides have given rise to both
optically bright and dark spin-valley configurations; the resulting optical
configuration is contingent on the lone electron's spin as well as its
occupation of either the outermost or innermost conduction bands in the $%
K^{\prime }$ valley \cite{Liu2019,NScRev2015}. It is important to note that
for intravalley $X^{-}$ trions with identical electron masses, the wave
function antisymmetrization is performed with respect to the spin states of
the two electrons. However, if the two intravalley electrons have unequal
masses, calculations must be performed in the context of a three-body
problem.

In contrast, the $X^{+}$ trion, consists of two holes in the outermost
valence band of the $K-$ and $K^{\prime }-$valleys as well as an unpaired
electron in either the innermost or outermost $K-$valley conduction
sub-bands. By the Pauli exclusion principle, the $X^{+}$ trion can only
exist in the intervalley configuration. It is important to note that for
intervalley $X^{+}$ trions, the wave function's antisymmetrization is
performed with respect to the spin states of the two holes that have
identical masses. Interestingly, spectroscopic measurements in MoS$_{2}$ and
MoSe$_{2}$ monolayers confirm the presence of $X^{+}$ spin-splitting in the
innermost and outermost $K-$valley conduction sub-bands, giving rise to both
optically bright and dark configurations. This is in stark contrast to the $%
X^{+}$ trions in monolayers such as WS$_{2}$ and WSe$_{2}$, neither of which
permits such overlapping of the $K-$ valley conduction sub-bands. Such
overlapping, which is uniquely characteristic of positive and negative
trions in MoS$_{2}$ and MoSe$_{2}$, is attributed to differences in the
effective masses of the conduction sub-bands caused by the spin-orbit
coupling \cite{Liu2019,NScRev2015}. \newline

\subsection{Binding energies in short- and long-range limits}

To calculate trion binding energies, one should solve the system of coupled
differential equations (\ref{TrionGeneral}) with the RK potential. The
latter system includes the electron and hole effective masses, a screening
distance, and dielectric constants of a supported or encapsulated material.
In our theoretical approach, all these characteristics are the input
parameters. In this sense, our approach has no fitting parameters. In our
calculations, we use the necessary parameters for the trions that were
calculated from first principles. The resulting binding energies of positive
or negative trions in the monolayer have a parametric dependence on the
screening distance $\rho _{0}$ and the electron, $m_{e}$, and hole, $m_{h}$,
effective masses. In our calculations we use the effective masses from Ref. 
\cite{Kormanyos2015} that are extracted from the low energy band structure
obtained in \textit{ab initio} density functional theory calculations to
describe the dispersion of the valence and conduction bands at their
extrema. The screening length was calculated using the polarizability $\chi $
for TMDCs given in Ref. \cite{Mosttop,Reichman2013}.

Before presenting the results for the trion binding energies obtained by the
numerical solution of the system of equations (\ref{TrionGeneral}) with the
potential\ (\ref{Keldysh}), we calculate the binding energies in both the
long- and short-range limits within the diagonal approximation for
freestanding TMDC\ monolayers. Calculations are performed using analytical
expressions (\ref{Colim}) and (\ref{Kelim}), respectively, for the single
state intravalley\ $X^{-}$\ trions formed in the $K$ valley, and are shown
in Table \ref{tab:limit}. For simplicity, in the binding energy calculation,
we consider the contribution of the term $K=0$. The latter leads to the
reduction of the equations in the diagonal approximation to only a single
equation.

\begin{table}[h]
\caption{ $X^{-}$ trion binding energies in meV in the short $r<<\protect%
\rho _{0}$ and long $r>>\protect\rho _{0}$ range limits. Calculations are
performed in the diagonal approximation, when $K=0$ for the state ($S,L$) =
(1/2,0) and with the electron and hole effective masses from Ref. 
\protect\cite{Kormanyos2015}. $m_{0}$ is the free electron mass. The
calculations are performed for freestanding monolayers. The data 
\protect\cite{Ganchev} are estimated using Fig. 1 from this work for the
ratio $m_e/m_h$ given in the Table.}
\label{tab:limit}
\begin{center}
\begin{tabular}{cccccccc}
\hline\hline
& \multicolumn{3}{c}{Input parameters} & $\text{Theory, }r<<\rho $ &  & $%
\text{Theory, }r>>\rho $ & $\text{Experiment}$ \\ 
& $m_{e}/m_{0}$ & $m_{h}/m_{0}$ & $\rho _{0}$, $\overset{o}{A}$ & This work
& \cite{Ganchev} & This work &  \\ \hline
$\text{MoS}_{2}$ & 0.350 & 0.428 & 38.62 \cite{Mosttop} & $27.2$ & $29-31$ & 
$21.1$ & $18\pm 1.5$ \cite{MoS23Heinz}, $30,32$ \cite{Zhang2014} \\ 
$\text{MoSe}_{2}$ & 0.38 & 0.44 & 51.71 \cite{Reichman2013} & $25.3$ & $%
29-31 $ & $21.3$ & $30$ \cite{MoSe21 Ross,MoSe2Singh} \\ 
$\text{WS}_{2}$ & 0.27 & 0.32 & 37.89 \cite{Reichman2013} & $29.1$ & $28-30$
& $22.5$ & $30$ \cite{WS2Plechinger}, $45$ \cite{bZu} \\ 
$\text{WSe}_{2}$ & 0.29 & 0.34 & 45.11 \cite{Reichman2013} & $27.5$ & $28-30$
& $21.8$ & $30$ \cite{WSe2 Jones,WSe2Wang} \\ \hline\hline
\end{tabular}%
\end{center}
\end{table}
The examination of the results in Table \ref{tab:limit} reveals that in the
short-range limit, $r\ll \rho _{0},$ the binding energies are less than
experimental values for MoSe$_{2}$, WS$_{2}$ and WSe$_{2}$. The binding
energy of $X^{-}$\ in MoSe$_{2}$ monolayer is also less than experimental
data reported in Ref. \cite{Zhang2014}. The same behaviors are observed for
the long-range limit, $r\gg \rho _{0}$. However, results with the
logarithmic potential are always larger than those with the Coulomb
potential. It is worth noting that the binding energies obtained for the
short-range limit are in reasonable agreement with calculations \cite%
{Ganchev}, in which trions were treated as three logarithmically interacting
particles. Thus, analytical expressions (\ref{Colim}) and (\ref{Kelim})
provide reasonable estimates for trion binding energies in TMDC monolayers.

\subsection{Binding energies of $X^{-}$ and $X^{+}$ trions}

While we can obtain the analytical expressions for the trion binding energy
and wave functions in both the long- and short-range limits within the
diagonal approximation, it is impossible to solve the trion problem
analytically when charged carriers interact via the centrally symmetric RK
potential. This conclusion follows based on the Nikiforov-Uvarov approach 
\cite{NikUvarov}: if there exists a class of transformations which allows us
to transform the three-body radial Schr\"{o}dinger equation with potential (%
\ref{Keldysh}) into an equation of the simpler form $F(r)\frac{d^{2}u(r)\ }{%
dr^{2}}+f(r)\frac{du(r)\ }{dr}+\xi u(r)=0$, where $F(r)$ and $f(r)$ are
polynomials of at most the second and the first degree, respectively, and $%
\xi $ is a constant, the corresponding equation can be solved analytically
within the theory of special functions. To the best of our knowledge no one
has found, at least by today, such a transformation for the three-body
radial Schr\"{o}dinger equation with RK potential (\ref{Keldysh}). Noticing
that the modified Kratzer potential considered in Ref. \cite{Molas2019} for
excitons description can be used to calculate the binding energy of trions.
For the modified Kratzer potential, follow \cite{KezFBS2023} and using the
approach presented for the long-range limit, one can obtain in the diagonal
approximation the analytical expression for the binding energy and wave
function for trions in terms of the Laguerre polynomials.

\begin{figure}[h]
\centering
\includegraphics[width=8.0cm]{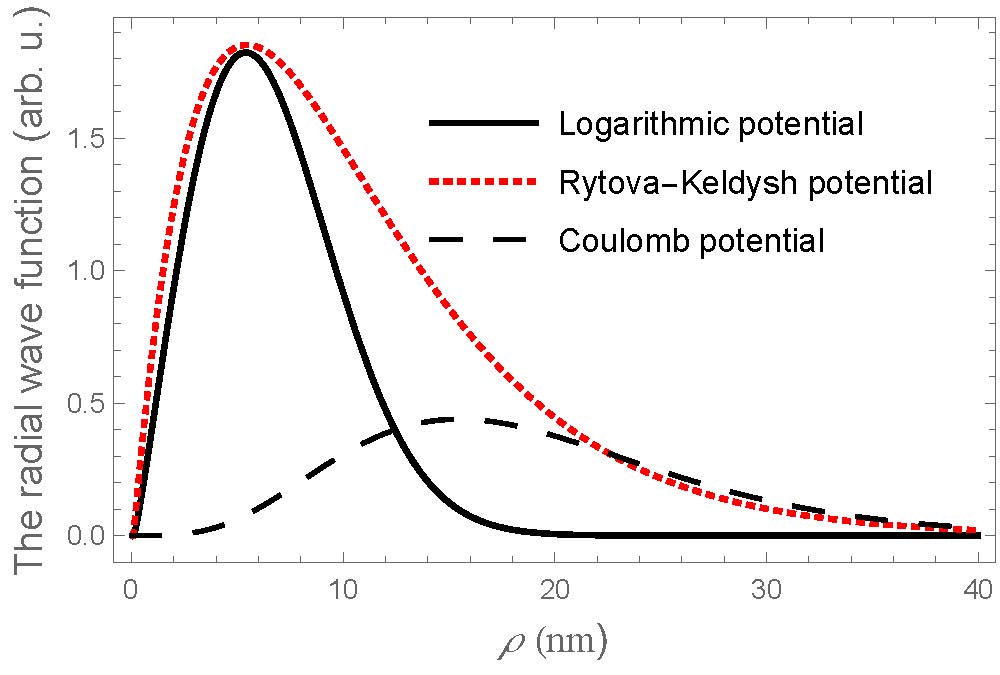}
\caption{(Color online) Dependence of the hyperradial wave function on
hyperradius $\protect\rho $ for the logarithmic, Coulomb, and Rytova-Keldysh
potentials. }
\label{Waverad}
\end{figure}

\begin{table}[h]
\caption{ Convergence of binding energies for the freestanding bright $X^{-}$
trion in meV. The electron and hole effective masses are taken from Ref. 
\protect\cite{Kormanyos2015}.}
\label{tab:convergion}
\begin{center}
\begin{tabular}{ccccccc}
\hline\hline
TMDC & $K=0$ & $K=0,2$ & $K=0,2,4$ & $K=0,2,4,6$ & $K=0,2,4,6,8$ & $%
K=0,2,4,6,8,10$ \\ \hline
& \multicolumn{6}{c}{Binding Energy, meV} \\ \hline
$\text{MoS}_{2}$ & 28.8 & 29.9 & 31.9 & 32.6 & 32.78 & 32.80 \\ 
$\text{MoSe}_{2}$ & 24.5 & 26.6 & 26.9 & 27.4 & 27.6 & 27.6 \\ 
$\text{WS}_{2}$ & 30.1 & 32.3 & 32.6 & 32.8 & 33.0 & 33.1 \\ 
$\text{WSe}_{2}$ & 24.5 & 25.5 & 27.2 & 27.8 & 28.2 & 28.3 \\ \hline\hline
\end{tabular}%
\end{center}
\end{table}
The system of coupled differential equations (\ref{TrionGeneral}) for the
hyperradial functions $u_{K\lambda }^{L}(\rho )$ are solved numerically. By
solving the system of equations (\ref{TrionGeneral}) one finds the binding
energy as well as the corresponding hyperradial functions. The latter allows
one to construct the wave function (\ref{ExpanTrion}). The convergence of
the binding energies for trions with respect to the grand angular momentum $%
K $ is given in Table \ref{tab:convergion}. The relative convergence is
checked as $\Delta B/B=\left[ B(K+2)-B(K)\right] /B(K)$, where $B(K)$ is the
binding energy for the given $K$. The analysis of the results in Table \ref%
{tab:convergion} shows that the reasonable convergence is reached for $%
K_{\max }$ = 12, so we limit our considerations to this value. The
dependence of hyperradial wave functions on hyperradius $\rho $ for a trion
obtained in the short-range limit (logarithmic potential) long-range limit
(the Coulomb potential), and for the Rytova-Keldysh potential are shown in
Fig. \ref{Waverad}.

\begin{figure}[h]
\centering
\includegraphics[width=8.0cm]{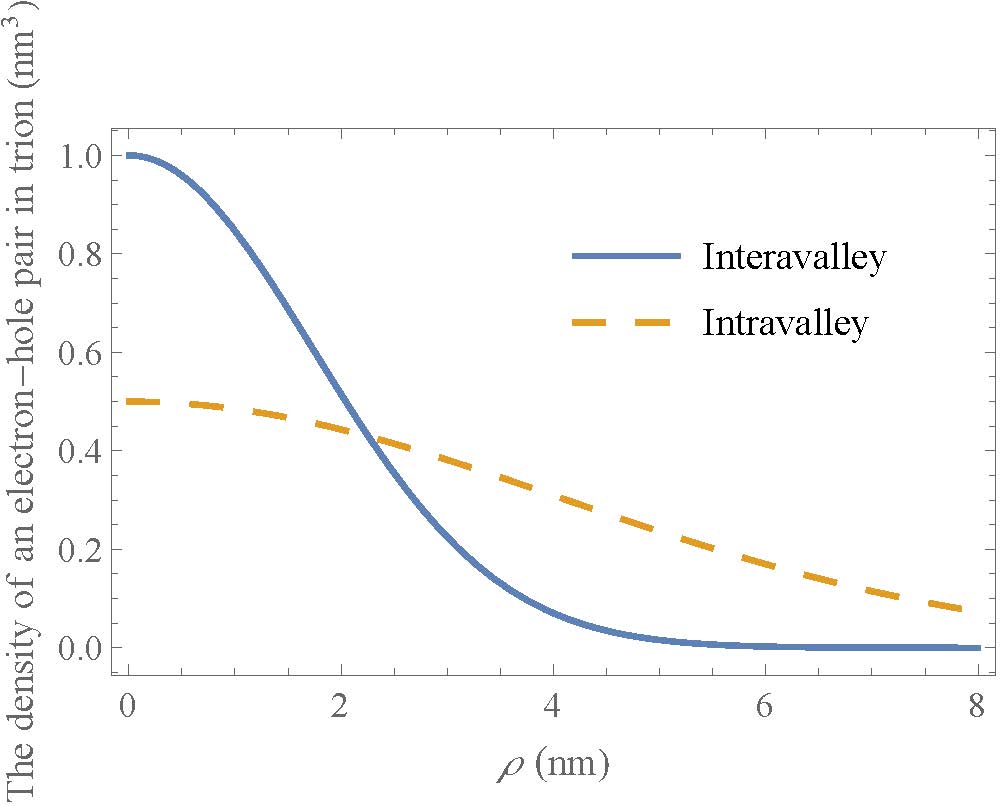}
\caption{(Color online) Probability density distribution for the intervalley
and intravalley electron-hole pair.}
\label{Density}
\end{figure}

The results of our calculations for the binding energies of the trions in
freestanding, supported by dielectric SiO$_{2}$, and encapsulated by hBN MoS$%
_{2}$, MoSe$_{2}$, WS$_{2}$, and WSe$_{2}$ monolayers, are presented in
Table \ref{tab:encapsulated}. The dielectric environment of an
atomically-thin monolayer of TMDC affects both the electronic band gap and
the excitonic binding energy in the monolayer. In our calculations, the role
of the substrate and the encapsulation by hBN is considered through the
potential (\ref{Keldysh}) with $\epsilon =(\varepsilon _{1}+\varepsilon
_{2})/2$, where $\varepsilon _{1}$ and $\varepsilon _{2}$ are the dielectric
constants of two materials that the TMDC layer is surrounded by. More
quantitative treatments of substrate and heterostructure screening effects
can be achieved with the quantum electrostatic heterostructure approach
developed in Refs. \cite{Latini2015,Thygesen2015}. In addition to the
intrinsic dielectric response of a monolayer TMDC and its electronic
environment, a number of other potential sources of screening should be
considered for a fully quantitative theory \cite{Latini2015,Florian2018}.
However, even in our simple approach one finds a strong dependence of the
binding energy on whether the TMDC monolayer is freestanding in air,
supported on SiO$_{2}$, or encapsulated in the hexagonal boron-nitride.

\begin{table}[h]
\caption{ $X^{-}$ and $X^{+}$ bright trions binding energies in meV in the
freestanding, SiO$_{2}$ - supported, and hBN - encapsulated TMDC monolayers.
The electron and hole effective masses are taken from Ref. \protect\cite%
{Kormanyos2015}. Effective masses of electrons from $K$ and $K^{\prime}$
valleys are the same. $m_{0}$ is the free electron mass.}
\label{tab:encapsulated}
\begin{center}
\begin{tabular}{cccccccc}
\hline\hline
& \multicolumn{3}{c}{Input parameters} & \multicolumn{2}{c}{Suspended} & 
Supported & Encapsulated \\ \hline
& $m_{e}/m_{0}$ & $m_{h}/m_{0}$ & $\rho _{0}$, $\overset{o}{A}$ & $X^{-}$ & $%
X^{+}$ & $X^{-}$ & $X^{-}$ \\ \hline
MoS$_{2}$ & 0.350 & 0.428 & 38.62 \cite{Mosttop} & 32.80 & 33.2 & 25.1 & 23.3
\\ 
MoSe$_{2}$ & 0.38 & 0.44 & 51.71 \cite{Reichman2013} & 27.6 & 28.8 & 20.2 & 
18.4 \\ 
WS$_{2}$ & 0.27 & 0.32 & 37.89 \cite{Reichman2013} & 33.1 & 33.2 & 25.6 & 
22.4 \\ 
WSe$_{2}$ & 0.29 & 0.34 & 45.11 \cite{Reichman2013} & 28.6 & 28.9 & 21.4 & 
18.2 \\ \hline\hline
\end{tabular}%
\end{center}
\end{table}

\begin{table}[h]
\caption{ The binding energy of intervalley $X^{-}$ and $X^{+}$
freestanding, supported by SiO$_{2}$ dielectric and encapsulated by hBN in
the monolayer WSe$_{2}$.}
\label{tab:WSe2}
\begin{center}
\begin{tabular}{cccc}
\hline\hline
\multicolumn{4}{c}{Binding energy of $X^{-}$ and $X^{+}$ trions in WSe$_{2}$%
, meV} \\ \hline
& Freestanding & Supported & Encapsulated \\ \hline
\multicolumn{4}{c}{Intravalley, singlet state, $L=0$, $S=1/2$} \\ \hline
$X^{-}$, $K$ valley electrons & 28.6 & 21.4 & 18.2 \\ 
$X^{-},$ $K^{\prime }$ valley electrons & 38.7 & 34.6 & 29.4 \\ 
\multicolumn{4}{c}{Intervalley, singlet state, $L=0$, $S=1/2$} \\ 
$X^{+}$, $K$ valley electron & 28.9 & 22.9 & 19.8 \\ 
$X^{+}$, $K^{\prime }$ valley electron & 29.4 & 24.5 & 21.6 \\ 
\multicolumn{4}{c}{Intervalley, triplet state, $L=1$, $S=3/2$} \\ 
$X^{-}$, $K,$ $K^{^{\prime }}$ valley electrons & 26.3 & 20.1 & 18.7 \\ 
\hline\hline
\end{tabular}%
\end{center}
\end{table}

\begin{figure}[h]
\centering
\includegraphics[width=7.0cm]{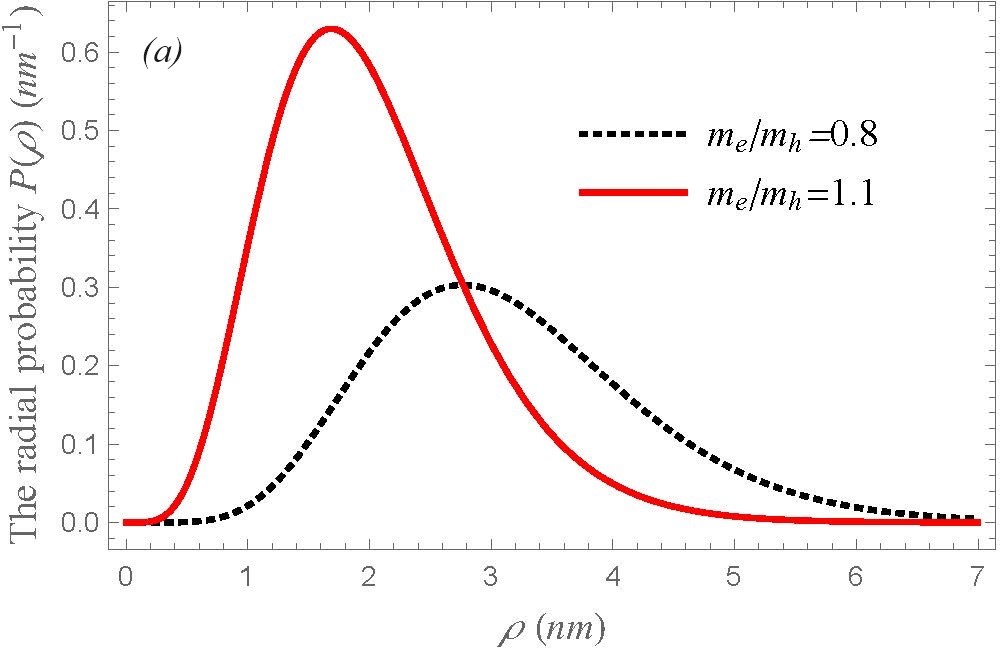} %
\includegraphics[width=9.5cm]{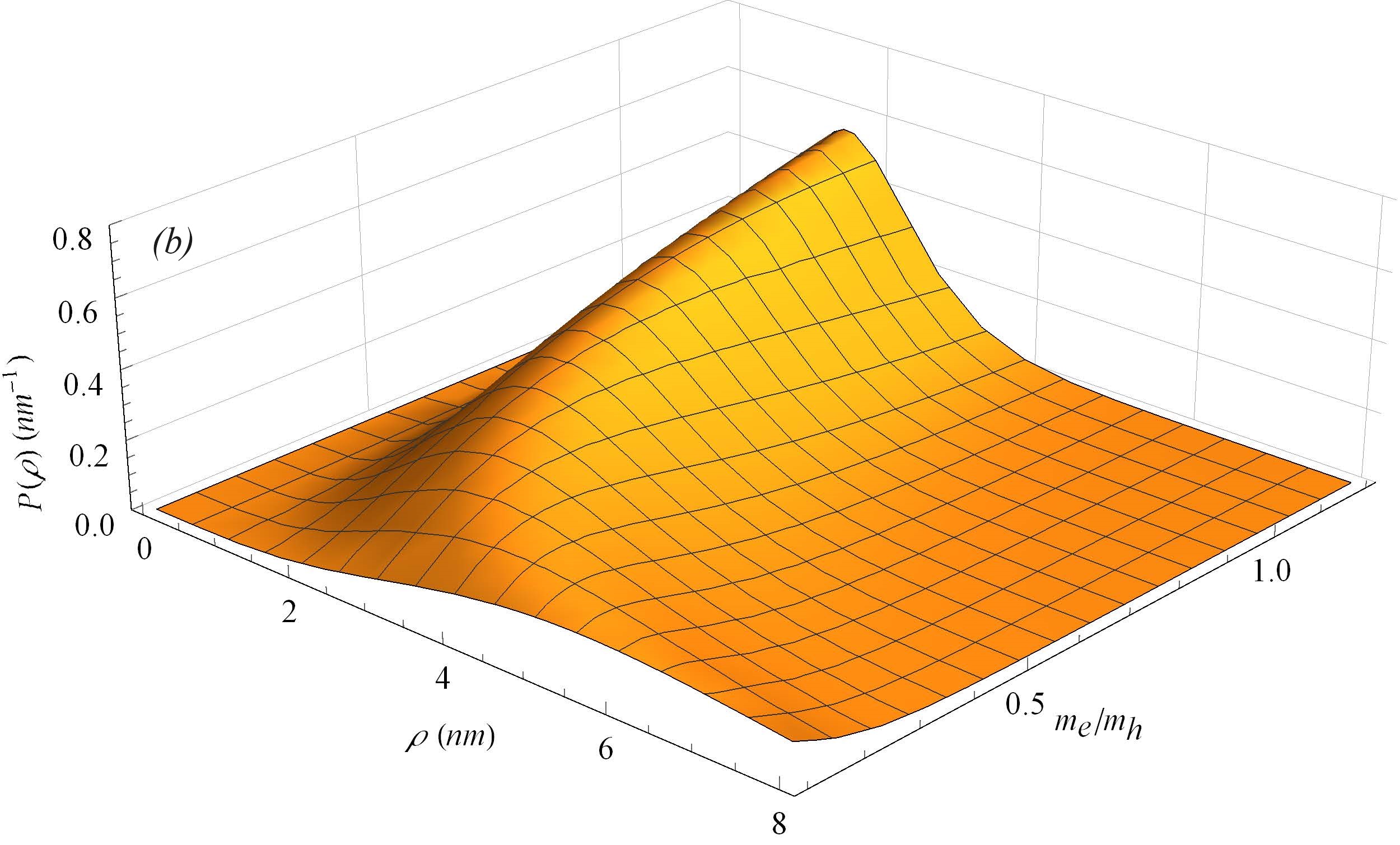} 
\includegraphics[width=5.6
cm]{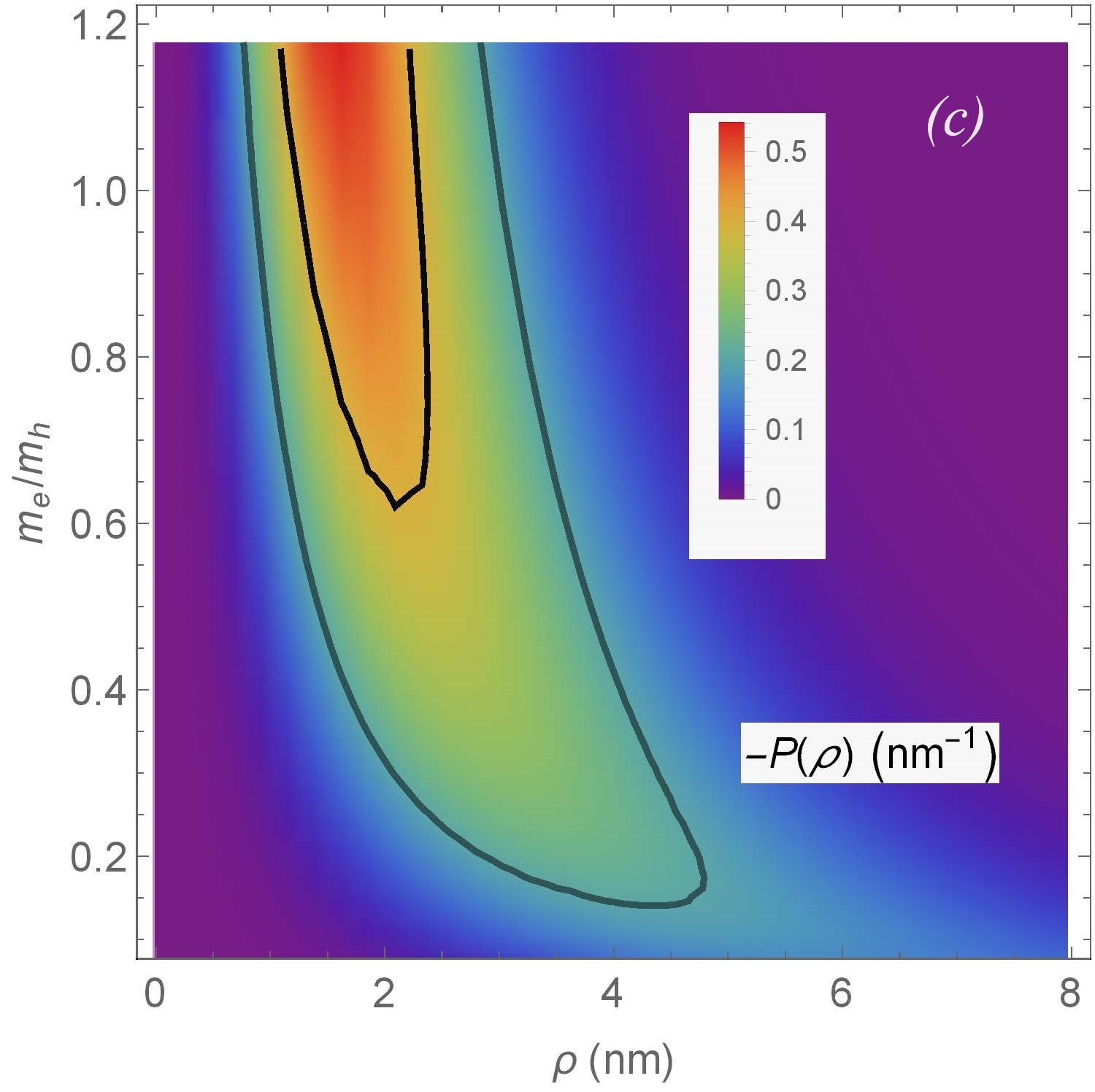}
\caption{(Color online) ($a$) Radial probability distribution for fixed $%
m_{e}/m_{h}$ ratio. ($b, c$) Dependence of the probability distribution of
three particles on hyperradius $\protect\rho$ and $m_{e}/m_{h}$ ratio. }
\label{Probep}
\end{figure}

The masses of electrons and holes in TMDC materials are of similar but not
the same magnitude. Following \textit{ab initio} calculations \cite%
{Kormanyos2015}, the effective masses of electrons and holes are different
in the $K$ and $K^{\prime }$ valleys. The effect of the different electron
masses in the $K$ and $K^{\prime }$ valleys on the binding energy of the
trion was studied in Ref. \cite{Dery2018}. At this point, intervalley $X^{-}$
in tungsten-based compounds is unique because its electrons have different
masses: one electron comes from the top spin-split of the conduction band of
one valley, while the second electron comes from the other valley \cite%
{Dery2016,Dery2015,CourtadeSemina}. The difference of electron masses breaks
the symmetry of the wave function and leads to the change in the binding
energy due to the following: the $X^{-}$ wave function is not antisymmetric
with respect to the permutation of two electrons; the binding energy
increases or decreases depending on the added excess electron mass. These
differences are consequential since the binding energy of the trion is
enhanced/suppressed when the added charge is heavier/lighter than the one
with the same charge in the neutral exciton (recall that the trion binding
energy is measured with respect to that of the exciton) \cite{Dery2018}. As
an example, we consider WSe$_{2}$ monolayer with the effective electron mass 
$m_{e}=0.29/m_{0}$ in one and $m_{e}=0.4/m_{0}$ in the other valley, while
the hole mass in the top valence band $m_{h}=0.34/m_{0}$ is less than the
mass of the added excess electron mass. The effect of different electron
masses from $K$ and $K^{\prime }$ valleys on the probability density
distribution for the intervalley and intravalley electron-hole pair is
demonstrated in Fig. \ref{Density}. We calculate a probability distribution
for three particles that form a trion. In Fig. \ref{Probep}$a$ interparticle
radial probability distributions for intravalley and intervalley $X^{-}$
trions are shown. The difference in the probability distribution is related
to the difference of the effective masses of electrons in $K$ and $%
K^{^{\prime }}$ and the symmetry of trion wave function in different
valleys: the singlet state $X^{-}$ trion (Fig. \ref{Probep}$a$, solid curve)
wave function is antisymmetric with respect to the permutation of two
electrons in the $K$ valley, while for the singlet state $X^{-}$ intervalley
trion (Fig. \ref{Probep}$a$, dotted curve) with different effective electron
masses in the $K$ and $K^{^{\prime }}$ valleys, the symmetry of the wave
function is broken. In Figs. \ref{Probep}$b$ and \ref{Probep}$c$ the
dependence of the probability distribution of three particles on the
hyperradius $\rho $ and the electron and hole masses ratio is shown. With
the increase of the ratio $m_{e}/m_{h}$, the particles become more compact
localized and, hence, more strongly bound.

We calculate the binding energies of trions when electrons in $K$ and $%
K^{^{\prime }}$ valleys have different effective masses. Let us consider
dark $X^{-}$ and $X^{+}$ trions in the spin-valley configuration shown in
Figs. \ref{TrionsValle}$e$ and \ref{TrionsValle}$h$, respectively. In Ref. 
\cite{Liu2019} observed both the positive and negative dark trions in WSe$%
_{2}$ under continuous electrostatic gating. The dark trions can be tuned
continuously between $X^{-}$ and $X^{+}$ trions with electrostatic gating 
\cite{Liu2019}. The bright trions have a higher energy in the range of
21--35 meV. Compared to bright trions, dark trions in WSe$_{2}$ have
smaller, but still sizable, binding energies 14--16 meV \cite{Liu2019}. The
authors reveal the spin-triplet configuration and distinct valley optical
emission of dark trions by their characteristic Zeeman splitting under a
magnetic field. Our calculations for the boron nitride-encapsulated $X^{-}$
spin-triplet configuration (Fig. \ref{TrionsValle}$e$) reveals the binding
energy 15.6 meV. For the dark spin-singlet $X^{+}$ (Fig. \ref{TrionsValle}$h$%
) we obtained 14.2 meV. These results match the theoretical prediction (15
meV) \cite{Deilmann2017} and the reported experimental value.

The analysis of binding energies will help us to understand the energy
difference between the binding energies of positive and negative trions in
molybdenum-based and tungsten-based TMDCs. As we mentioned above, in the
case when electrons are from different valleys, the wave function of the
intravalley $X^{-}$ trion has no symmetry and should not be antisymmetric
with respect to the permutation of two electrons. Results of calculations
for the binding energy of freestanding, SiO$_{2}$-supported, and
hBN-encapsulated monolayer WSe$_{2}$ intervalley $X^{-}$ and $X^{+}$ are
presented in Table \ref{tab:WSe2}. For a comparison, the same Table presents
binding energies for the intravalley $X^{-}$ and intervalley $X^{+}$ trions
formed with electrons from the $K$ and $K^{^{\prime }}$ valleys,
respectively. The analysis of the results leads to the following
conclusions: i. the binding energies of $X^{-}$ and $X^{+}$ trions with
electrons from the $K$ valley are always smaller; ii. the binding energy of
the $X^{-}$ trion in the state $L=1$ and $S=3/2$ with electrons from the $K$
and $K^{^{\prime }}$ valleys is the smallest; iii. the screening
substantially decreases the binding energies of both the $X^{-}$ and $X^{+}$
trions. To understand these patterns, for simplicity, consider the
analytical expression for the ground state energy (\ref{Colim}) in the
diagonal approximation, which allows us to obtain a physically transparent
picture of the dependence of the trion binding energy on the effective
masses of constituent quasiparticles. One can see the explicit dependence of
the energy on the three-particle effective mass $\mu$ given by Eq. (\ref%
{Colim}). The increase of $\mu$ leads to the increase of the binding energy
of the trion. The larger binding energies of $X^{-}$ and $X^{+}$ trions
formed with electrons from the $K^{^{\prime }}$ valley is due to the larger
their three-particle effective mass. The decrease of the $X^{-} $ binding
energy in the state $L=1$ and $S=3/2$ is due to the centrifugal repulsion
and different three-particle effective mass obtained with the masses of $K$
and $K^{^{\prime }}$ valley electrons.

Results of our calculations show the strong correlation between binding
energies of $X^{-}$ and $X^{+}$ trions and the constituent masses: more
massive holes correspond to larger binding energies of $X^{+}$ trions. The
binding energy of intervalley $X^{-}$ is larger than that of $X^{+}$ trions
if the mass of one of the valley electrons $m_{e}>m_{h}$.

\section{Conclusions}

We develop the theoretical formalism and study the formation of valley
trions in two-dimensional materials within the framework of a
non-relativistic potential model using the method of hyperspherical
harmonics in four-dimensional space. We solve the three-body Schr\"{o}dinger
equation with the Rytova-Keldysh potential by expanding the wave functions
of a trion in terms of the HH. The antisymmetrization of the wave functions
of $X^{-}$ and $X^{+}$ trions is based on the electron and hole spins and
valley indices, in case of two identical electrons or two identical holes.
In the case of $X^{-}$, when the two intravalley electrons could have
unequal masses, calculations are performed by solving the Schr\"{o}dinger
equation for three indistinguishable particles. We applied our theoretical
approach to trions in TMDC and demonstrated that the proposed theory is
capable of describing trions in TMDCs

We considered the long-range approximation when the RK potential can be
approximated by the Coulomb potential and the short-range limit when this
potential is approximated by the logarithmic potential. In the diagonal
approximation in both limits, the system of differential equations for the
hyperradial functions is decoupled. Our approach yields the analytical
solution for binding energies of trions in the diagonal approximation for
these two limiting cases - the Coulomb and logarithmic potentials. We
obtained the exact analytical expressions for eigenvalues and eigenfunctions
for $X $ $^{-}$ and $X^{+}$ trions. The corresponding energies can be
considered as the lower and upper limits for the trion binding energies.

The system of coupled differential equations 
for the hyperradial functions $u_{K\lambda }^{L}(\rho )$ can be solved
efficiently with fairly accurate results. In our approach, results for the
binding energies depend on input parameters and spin-valley symmetry
description of $X$ $^{-}$ and $X^{+}$ trions. On one hand, the difference in
binding energy between $X^{+}$ and $X^{-}$ is related to the effective
masses of the electrons and hole as well as screening distance for TMDC
monolayers. On the other hand, spin-valley quantum numbers combination. The
latter leads to the existence of bright and dark trions. The binding
energies of dark trions are less than binding energies of the bright trions. 

Let us note that different methods of calculations of the electron and hole
masses give slightly different effective masses. Thus, in our approach,
where the effective electron and hole masses are input parameters, we show
that trion binding energies are sensitive to these parameters and depend on
the wave function symmetry. Results of our calculations show the strong
correlation between binding energies of $X^{-}$ and $X^{+}$ trions and the
electron and hole masses. The more massive holes correspond to larger
binding energies of $X^{+}$ trions. When the mass of one of electrons form
one valley $m_{e}>m_{h}$, the binding energy of intervalley $X^{-}$ is
larger than that of $X^{+}$. Results of numerical calculations for the
ground state energies are in good agreement with similar calculations for
the Rytova-Keldysh potential and in fair agreement with the reported
experimental measurements for trion binding energies.

As mentioned in the introduction, there have been many other methods to
calculate the trion binding energy. All these methods are quite powerful to
study trions and give close results for binding energies, which are in
reasonable agreement with experimental measurements. The HH formalism yields
the analytical solution for binding energy and wave function of trions in
the diagonal approximation for the Coulomb and logarithmic potentials,
allows direct access to the wave functions, and explicit analysis of the
dependence of $X^{-}$ and $X^{+}$ binding energies on the effective masses
of constituent particles.

\appendix

\section{Evaluation of the $\mathcal{G}_{K\protect\lambda K^{^{\prime }}%
\protect\lambda ^{^{\prime }}}^{L}$for the Coulomb potential}

For the Coulomb potential the $\mathcal{G}_{K\lambda K^{^{\prime }}\lambda
^{^{\prime }}}^{L}$ can be evaluated analytically. Let us write (\ref%
{gCoulomb}) using the corresponding set of $\Omega _{i}\equiv (\alpha _{i},%
\widehat{\mathbf{x}}_{i},\widehat{\mathbf{y}}_{i})\ $angles

\begin{eqnarray}
\mathcal{G}_{K\lambda K^{^{\prime }}\lambda ^{^{\prime }}}^{L}
&=&ke^{2}\left\{ b_{1}\int \Phi _{K\lambda }^{L\ast }(\Omega _{i})\left( 
\frac{1}{\cos \alpha _{1}}\right) \Phi _{K^{^{\prime }}\lambda ^{^{\prime
}}}^{L^{^{\prime }}}(\Omega _{1})d\Omega _{1}\right. -  \notag \\
&&b_{2}\sum\limits_{\lambda _{k}\lambda _{k}^{^{\prime }}}\left\langle
\lambda _{k}\right\vert \lambda _{i}\rangle _{KL}\left\langle \lambda
_{k}^{^{\prime }}\right\vert \lambda _{i}\rangle _{KL}\int \Phi _{K\lambda
}^{L\ast }(\Omega _{2})\left( \frac{1}{\cos \alpha _{2}}\right) \Phi
_{K^{^{\prime }}\lambda ^{^{\prime }}}^{L^{^{\prime }}}(\Omega _{2})d\Omega
_{2}-  \notag \\
&&\left. b_{3}\sum\limits_{\lambda _{j}\lambda _{j}^{^{\prime
}}}\left\langle \lambda _{j}\right\vert \lambda _{i}\rangle
_{KL}\left\langle \lambda _{j}^{^{\prime }}\right\vert \lambda _{i}\rangle
_{KL}\int \Phi _{K\lambda }^{L\ast }(\Omega _{3})\left( \frac{1}{\cos \alpha
_{3}}\right) \Phi _{K^{^{\prime }}\lambda ^{^{\prime }}}^{L^{^{\prime
}}}(\Omega _{3})d\Omega _{3}\right\}  \label{Wcoulomb}
\end{eqnarray}%
The values of all integrals in (\ref{Wcoulomb}) are the same and for the
integral (\ref{CoulombIntergral}) we finally obtain

\begin{eqnarray}
\mathcal{C}_{K\lambda K^{^{\prime }}\lambda ^{^{\prime }}}^{(i)} &=&\int
\Phi _{K\lambda }^{L\ast }(\Omega )\left( \frac{1}{\cos \alpha }\right) \Phi
_{K^{^{\prime }}\lambda ^{^{\prime }}}^{L^{^{\prime }}}(\Omega )d\Omega
=\int \Phi _{K\lambda }^{L\ast }(\Omega )\left( \frac{1}{\cos \alpha }%
\right) \Phi _{K^{^{\prime }}\lambda ^{^{\prime }}}^{L^{^{\prime }}}(\Omega
)\cos \alpha \sin \alpha d\alpha d\varphi _{x}d\varphi _{y}=  \notag \\
&&\frac{1}{2}N_{K}^{\{l\}}N_{K^{^{\prime }}}^{\{l^{^{\prime }}\}}\delta
_{l_{1}l_{2}}\delta _{l_{1}^{^{\prime }}l_{2}^{^{\prime }}}\delta
_{LL^{\prime }}\delta _{MM^{\prime }}\Gamma (n+l_{1}+1/2)\Gamma
(n+l_{2}+1/2)\times   \notag \\
&&\Gamma (n^{\prime }+l_{1}+1/2)\Gamma (n^{\prime }+l_{2}+1/2)\times   \notag
\\
&&\sum_{\sigma =0}^{n}\sum_{\sigma ^{\prime }=0}^{n^{\prime }}\left\{ \frac{%
(-1)^{n-\delta }}{\Gamma (\sigma +1/2)\Gamma (\sigma +l_{1}+1/2)\Gamma
(n-\sigma +l_{2}+1/2)\Gamma (n-\sigma +1)}\right. \times   \notag \\
&&\frac{(-1)^{n^{\prime }-\delta ^{\prime }}}{\Gamma (\sigma ^{\prime
}+1/2)\Gamma (\sigma ^{\prime }+l_{1}+1/2)\Gamma (n^{\prime }-\sigma
^{\prime }+l_{2}+1/2)\Gamma (n^{\prime }-\sigma ^{\prime }+1)}\times   \notag
\\
&&\left. B[l_{1}+\sigma +\sigma ^{\prime }+1,l_{2}+(n+n^{\prime })-(\sigma
+\sigma ^{\prime })+1/2]\right\} ,
\end{eqnarray}%
with $\Gamma $ and $B\ $indicating the standard Gamma and Beta functions 
\cite{Gradshteyn,Abramowitz}, respectively, $n=(K-l_{1}-l_{2})/2$, and $%
n^{\prime }=(K^{\prime }-l_{1}^{\prime }-l_{2}^{\prime })/2.$

\section{Evaluation of the $\mathcal{L}_{K\protect\lambda K^{^{\prime }}%
\protect\lambda ^{^{\prime }}}^{(i)}$ for the logarithmic potential}

Let us consider the evaluation of \ $\mathcal{L}_{K\lambda K^{^{\prime
}}\lambda ^{^{\prime }}}^{(i)}$ defined in Eq. {(\ref{LKK}) }

\begin{eqnarray}
\mathcal{L}_{K\lambda K^{^{\prime }}\lambda ^{^{\prime }}}^{(i)} &=&\int
\Phi _{K\lambda }^{L\ast }(\Omega _{i})\ln \cos \alpha \Phi _{K^{^{\prime
}}\lambda ^{^{\prime }}}^{L^{^{\prime }}}(\Omega )\text{{}}d\Omega =\int
\Phi _{K\lambda }^{L\ast }(\Omega )\ln \cos \alpha \Phi _{K^{^{\prime
}}\lambda ^{^{\prime }}}^{L^{^{\prime }}}(\Omega )\cos \alpha \sin \alpha
d\alpha d\varphi _{x}d\varphi _{y}=  \notag \\
&&\frac{1}{4}N_{K}^{\{l\}}N_{K^{^{\prime }}}^{\{l^{^{\prime }}\}}\delta
_{l_{1}l_{2}}\delta _{l_{1}^{^{\prime }}l_{2}^{^{\prime }}}\delta
_{LL^{\prime }}\delta _{MM^{\prime }}\sum_{\sigma =0}^{n}\sum_{\sigma
^{\prime }=0}^{n^{\prime }}\left\{ \binom{\sigma +l_{1}}{\sigma }\binom{%
\sigma +l_{2}}{n-\sigma }\binom{\sigma ^{\prime }+l_{1}^{\prime }}{\sigma
^{\prime }}\binom{\sigma +l_{2}^{\prime }}{n^{\prime }-\sigma ^{\prime }}%
\times \right.   \notag \\
&&H[1/2\sigma l_{1}+l_{1}^{\prime }+\sigma ^{\prime }]-H[1/2(3-2\sigma
+2n+\sigma l_{1}+l_{2}+l_{1}^{\prime }-2(-1+l_{2}^{\prime }\sigma ^{\prime
}+2l_{2}^{\prime }n^{\prime }]\times   \notag \\
&&\left. \frac{\Gamma (\sigma +\sigma l_{1}+1/2l_{1}^{\prime }+\sigma
^{\prime })\Gamma (3/2-\sigma +n+1/2l_{2}+l_{2}^{\prime }(n^{\prime }-\sigma
^{\prime })}{\Gamma (2\sigma ^{\prime }+n+1/2l_{2}+\sigma
l_{1}+1/2l_{1}^{\prime }+l_{2}^{\prime }(\sigma -n^{\prime })+3/2))}\right\}
,
\end{eqnarray}%
where \ $H[z]$ is a Harmonic number.

\section{Reduction of Eq. (\protect\ref{RewEq}) to the Weber's equation}

The logarithmic potential is widely used as a quark confinement potential in
high energy physics with applications to elementary particle spectroscopy.
There is no known complete analytical solution of the Schr\"{o}dinger
equation for a logarithmic potential. However, there is known an almost
complete solution of the Schr\"{o}dinger equation for a logarithmic
potential \cite{Gesztesy1978,Muller1979,Khelashvili1}. Following the
methodology \cite{Muller1979} by the rescaling of variable $\rho $ and
introducing a new function, one can reduce (\ref{RewEq}) to the equation for
the parabolic cylinder function known Weber's equation \cite{Weber 1869}.
Let us rescale variable $\rho $ by introducing the new variable $z$ as 
\begin{equation}
\rho =e^{z-\alpha /\beta },\text{ }(-\infty <z<\infty )  \label{Z variable}
\end{equation}%
and set the solution of Eq. (\ref{RewEq}) in the form

\begin{equation}
u_{K\lambda }=e^{z-\alpha /\beta }\Phi (z).  \label{Function F}
\end{equation}%
Using (\ref{Z variable}) and (\ref{Function F}) one can rewrite (\ref{RewEq}%
) in the following form 
\begin{equation}
\frac{d^{2}\Phi (z)}{dZ^{2}}+[-\Delta ^{2}+V(z)]\Phi (z)=0,
\label{Main Equation}
\end{equation}%
where $\Delta ^{2}=(K+1)^{2}$ and 
\begin{equation}
V(z)=-z\beta e^{2\alpha /\beta +2z}.
\end{equation}%
The function $V(z)$ is twice differentiable at a stationary point $z_{0}$.
Finding the existence of eigenvalues for differential equations is a
boundary value problem: $-\Delta ^{2}+V(z)$ must be positive and in the
vicinity of the maximum value of $V(z)$, where $-\Delta ^{2}+V(z)>0$ Eq. (%
\ref{Main Equation}) leads to an oscillatory type solution. Let us find the
value of $z=z_{0}$ for which $V(z)$ has a maximum. In the vicinity of this
maximum $-\Delta ^{2}+V(z)$ can become positive. The condition for the
derivative $\frac{dV(z)}{dz}=0$ and $\left. \frac{d^{2}V(z)}{dz^{2}}%
\right\vert _{z=z_{0}}<0$ yields the value $z_{0}=-\frac{1}{2}.$ Expanding $%
V(z)$ in a Taylor series in the neighborhood of the maximum at $z_{0}$, we
obtain in the first order with respect to $(z-z_{0})^{2}$

\begin{equation}
V(z)=V(z_{0})+\frac{z-z_{0}}{2!}V^{\prime \prime }(z_{0})+...,
\label{Expansion V(z)}
\end{equation}%
where 
\begin{equation}
V^{\prime \prime }(z_{0})=-4\beta e^{^{2\alpha /\beta +2z_{0}}}(1+z_{0}).
\end{equation}

Let's set the new variable 
\begin{equation}
w=S(z-z_{0}\ ),\text{ \ \ where \ }S=\sqrt[4]{4\beta e^{^{(2\alpha -\beta
)/\beta }}}.  \label{w variable}
\end{equation}%
Applying (\ref{w variable}) to (\ref{Main Equation}), we have 
\begin{equation}
\left[ \frac{d^{2}\Phi (z)}{dw^{2}}+\frac{-\Delta ^{2}+1/8S^{4}}{S^{2}}-%
\frac{1}{4}w^{2}\right] \Phi (w)=\overset{\infty }{\sum }_{i=3}\frac{%
(i-1)2^{i-3}}{i!}\frac{w^{i}}{s^{i-2}}\Phi (w).
\end{equation}%
Neglecting the right side in the latter equation, we obtain 
\begin{equation}
\left[ \frac{d^{2}\Phi (z)}{dw^{2}}+\frac{-\Delta ^{2}+1/8S^{4}}{S^{2}}-%
\frac{1}{4}w^{2}\right] \Phi (w)=0.  \label{CylinderEq}
\end{equation}%
The eigenvalues and eigenfunctions for the a boundary value problem for Eq. (%
\ref{CylinderEq}) with the vanishing wave function at infinity can be
obtained by comparing this equation with the equation for the parabolic
cylinder functions. This comparison requires that the solution of (\ref%
{CylinderEq}) be square-integrable only if 
\begin{equation}
\frac{-\Delta ^{2}+1/8S^{4}}{S^{2}}=\frac{(2n+1)}{2},\text{ \ }n=0,1,2...
\label{S1}
\end{equation}%
Equation (\ref{S1}) can be reduced to the following biquadratic equation 
\begin{equation}
S^{4}-4(2n+1)S^{2}-8\Delta ^{2}=0,\text{ \ }n=0,1,2...
\end{equation}%
with solution 
\begin{equation}
S^{2}=2(2n+1)+2\left[ (2n+1)^{2}+2\Delta ^{2}\right] ^{1/2}.  \label{S2Eq}
\end{equation}%
After substituting (\ref{w variable}) into Eq. (\ref{S2Eq}), we get 
\begin{equation}
2\beta e^{^{(2\alpha -\beta )/2\beta }}=4\left( (2n+1)+\left[
(2n+1)^{2}+2\Delta ^{2}\right] ^{1/2}\right) ^{2}.
\end{equation}%
Therefore, 
\begin{equation}
\beta e^{^{(2\alpha -\beta )/2\beta }}=2\left( (2n+1)+\left[
(2n+1)^{2}+2\Delta ^{2}\right] ^{1/2}\right) ^{2},
\end{equation}%
\begin{equation}
\text{ln}\beta +\text{ln}e^{^{(2\alpha -\beta )/2\beta }}=2\text{ln}2\left(
(2n+1)+\left[ (2n+1)^{2}+2\Delta ^{2}\right] ^{1/2}\right)
\end{equation}%
or finally 
\begin{equation}
\alpha =\text{ln}2\left( (2n+1)+\left[ (2n+1)^{2}+2\Delta ^{2}\right]
^{1/2}\right) -\frac{\text{ln}\beta +\beta }{2}.
\end{equation}%
Thus, 
\begin{equation}
\kappa ^{2}=\text{ln}2\left( (2n+1)+\left[ (2n+1)^{2}+2\Delta ^{2}\right]
^{1/2}\right) -\frac{\text{ln}\beta +\beta }{2}+\frac{2\mu }{\hbar ^{2}}%
\left( -\frac{ke^{2}}{\epsilon \rho _{0\text{ }}}\ln \rho _{0\text{ }}+%
\mathcal{B}_{123}+\mathcal{J}_{K\lambda K\lambda }\right) .
\end{equation}%
The eigenfunctions that correspond to these eigenvalues are 
\begin{equation}
\Phi =D_{n}(w)=2^{(n-1)/2}e^{-w^{2}/4}F\left( \frac{1-n}{2},\frac{3}{2},%
\frac{w^{2}}{2}\right) ,  \label{DCFunc}
\end{equation}%
where $w$ is defined by Eq. (\ref{w variable}) and $F(\frac{1-n}{2},\frac{3}{%
2},\frac{w^{2}}{2})$ is a confluent hypergeometric function.

\end{document}